\magnification=\magstep1 \overfullrule=0pt 
\advance\hoffset by -0.35truecm   
\vsize=23.1truecm
\font\tenmsb=msbm10       \font\sevenmsb=msbm7
\font\fivemsb=msbm5       \newfam\msbfam
\textfont\msbfam=\tenmsb  \scriptfont\msbfam=\sevenmsb
\scriptscriptfont\msbfam=\fivemsb
\def\Bbb#1{{\fam\msbfam\relax#1}}
\def\R{{\Bbb R}}\def\Z{{\Bbb Z}}

\def\hh{{\Bbb H}}

\font\gros=cmssbx10 scaled \magstep2
\font\kap=cmcsc10     

\font\sf=cmss10
\font\fat=cmmib10 
\font\smfat=cmmib7
\font\large=cmbx9 scaled \magstep2
\def\lb{\lbrack}\def\rb{\rbrack}  \def\q#1{$\lb${\rm #1}$\rb$}
\def\bn{\bigskip\noindent}\def\mn{\medskip\smallskip\noindent}
\def\sn{\smallskip\noindent} 

\def\cedille#1{\setbox0=\hbox{#1}\ifdim\ht0=1ex \accent'30 #1%
 \else{\ooalign{\hidewidth\char'30\hidewidth\crcr\unbox0}}\fi}

\def\qed{{\vrule height4pt width4pt depth1pt}}

\def\hbn{\hfill\break\noindent}
\def\sqr#1#2{{\vcenter{\vbox{\hrule height.#2pt
       \hbox{\vrule width.#2pt height #1pt \kern#1pt 
         \vrule width.#2pt}\hrule  height.#2pt}}}}

\global\newcount\glgnum
\global\glgnum=0
\def\glg{{{\global\advance\glgnum by1}{(\number\glgnum)}}}
\def\mkg#1{\glg\xdef#1{(\the\glgnum)}}
\global\newcount\refnum
\global\refnum=0
\def\preref{{\global\advance\refnum by1}{\q{\number\refnum}}}
\def\ref#1{\preref\xdef#1{\the\refnum}}
\def\eq{\eqno\mkg} 
\def\mkgho#1{\xdef#1{(\the\glgnum}}
\def\mkgvo#1{\xdef#1{\the\glgnum)}}
\def\mkgvho#1{\xdef#1{\the\glgnum}}
\def\bcr{\cr & & \cr}  
\def\be{$$}    \def\ee{$$}  \def\labela#1{\eq#1}
\def\ba{$$ \eqalignno }      \def\ea{$$}
\def\frac#1#2{{#1 \over #2}} 
\def\labelb#1{&\mkg#1}
\def\refe#1{{#1}}
\def\section#1{\bn\bn{\large #1}\bn}
\def\subsection#1{\bn{\bf #1}.\ \ }
\def\mbox#1{ {\rm #1} } 
\def\sc{\it}
\def\lefteqn#1{& #1}
\def\hspace#1{\phantom{#1}}
\def\ja{{a}}  \def\s{\sigma}
\def\rg{{\rm g}} 
\def\bfz{{\vec z}}
 
\def\bfx{{\vec x}}
\def\jba{{\overline a}}
\def\cH{{\cal H}}
\def\cW{{\cal W}} 
\def\bL{{\overline L}}
\def\cite#1{\ref{#1}}  
\def\d{\delta} 
\def\bJ{{\overline J}}
\def\bT{{\overline T}}
\def\bX{\overline X}  
\def\bz{\bar z} 
\def\bw{\bar w}
\def\Im{{\rm Im}} 
\def\Re{{\rm Re}} 
\def\Res{{\rm Res}}

\def\mZh{{\Z + \frac12}}
\def\mZ{{\Z}}
\def\H{{\cal H}}

\def\u{\upsilon}
\def\c{\gamma}

\def\vac{{\,|\sigma\rangle}}

\def\cav{{\langle \sigma |\,}}
\def\nn{\ }
\def\cJ{\hbox{{\sf J}}}
\def\cT{\hbox{{\sf T}}}
\def\p{\partial}
\def\bp{\bar \partial}  
\def\o{w} \def\u{u}  
\def\ome{\omega} 
\def\oh{{1\over2}}
\def\no{\left(\frac{i}{2}\right)^h}
\def\noq{\left(\frac{i}{2}\right)^{2h}}
\def\der{{\rm x}}
\def\ter{{\rm y}}
\def\up{^{{\scriptscriptstyle (P)}}}\def\uh{^{{\scriptscriptstyle (H)}}}
\newread\epsffilein    \newif\ifepsffileok    \newif\ifepsfbbfound   
\newif\ifepsfverbose   \newdimen\epsfxsize    \newdimen\epsfysize    
\newdimen\epsftsize    \newdimen\epsfrsize    \newdimen\epsftmp      
\newdimen\pspoints  \pspoints=1bp  \epsfxsize=0pt  \epsfysize=0pt         
\def\epsfbox#1{\global\def\epsfllx{72}\global\def\epsflly{72}%
 \global\def\epsfurx{540}\global\def\epsfury{720}%
 \def\lbracket{[}\def\testit{#1}\ifx\testit\lbracket
 \let\next=\epsfgetlitbb\else\let\next=\epsfnormal\fi\next{#1}}%
\def\epsfgetlitbb#1#2 #3 #4 #5]#6{\epsfgrab #2 #3 #4 #5 .\\%
 \epsfsetgraph{#6}}%
\def\epsfnormal#1{\epsfgetbb{#1}\epsfsetgraph{#1}}%
\def\epsfgetbb#1{\openin\epsffilein=#1
\ifeof\epsffilein\errmessage{I couldn't open #1, will ignore it}\else
 {\epsffileoktrue \chardef\other=12
  \def\do##1{\catcode`##1=\other}\dospecials \catcode`\ =10 \loop
  \read\epsffilein to \epsffileline \ifeof\epsffilein\epsffileokfalse\else
       \expandafter\epsfaux\epsffileline:. \\    \fi
  \ifepsffileok\repeat   \ifepsfbbfound\else
  \ifepsfverbose\message{No bounding box comment in #1; using defaults}\fi\fi
  }\closein\epsffilein\fi}%
\def\epsfclipstring{}%

\def\epsfsetgraph#1{ \epsfrsize=\epsfury\pspoints 
 \advance\epsfrsize by-\epsflly\pspoints  \epsftsize=\epsfurx\pspoints
 \advance\epsftsize by-\epsfllx\pspoints \epsfxsize\epsfsize\epsftsize
  \epsfrsize
\ifnum\epsfxsize=0\ifnum\epsfysize=0\epsfxsize=\epsftsize\epsfysize=\epsfrsize
   \epsfrsize=0pt \else\epsftmp=\epsftsize \divide\epsftmp\epsfrsize
  \epsfxsize=\epsfysize \multiply\epsfxsize\epsftmp
  \multiply\epsftmp\epsfrsize \advance\epsftsize-\epsftmp \epsftmp=\epsfysize
  \loop \advance\epsftsize\epsftsize \divide\epsftmp 2 \ifnum\epsftmp>0
  \ifnum\epsftsize<\epsfrsize\else \advance\epsftsize-\epsfrsize
  \advance\epsfxsize\epsftmp \fi \repeat \epsfrsize=0pt \fi \else
  \ifnum\epsfysize=0 \epsftmp=\epsfrsize \divide\epsftmp\epsftsize
  \epsfysize=\epsfxsize \multiply\epsfysize\epsftmp
  \multiply\epsftmp\epsftsize \advance\epsfrsize-\epsftmp \epsftmp=\epsfxsize
  \loop \advance\epsfrsize\epsfrsize \divide\epsftmp 2 \ifnum\epsftmp>0
  \ifnum\epsfrsize<\epsftsize\else \advance\epsfrsize-\epsftsize
  \advance\epsfysize\epsftmp \fi \repeat \epsfrsize=0pt \else
  \epsfrsize=\epsfysize \fi \fi \ifepsfverbose\message{#1:
  width=\the\epsfxsize, height=\the\epsfysize}\fi \epsftmp=10\epsfxsize
  \divide\epsftmp\pspoints \vbox to\epsfysize{\vfil\hbox to\epsfxsize{
  \ifnum\epsfrsize=0\relax \includegraphics{#1} \else \epsfrsize=10\epsfysize \divide\epsfrsize\pspoints
  \includegraphics{#1}\fi \hfil}}%
\global\epsfxsize=0pt\global\epsfysize=0pt}%
{\catcode`\%=12 \global\let\epsfpercent=
\long\def\epsfaux#1#2:#3\\{\ifx#1\epsfpercent
   \def\testit{#2}\ifx\testit\epsfbblit    \epsfgrab #3 . . . \\%
      \epsffileokfalse     \global\epsfbbfoundtrue
   \fi\else\ifx#1\par\else\epsffileokfalse\fi\fi}%
\def\epsfempty{}\def\epsfgrab #1 #2 #3 #4 #5\\{%
\global\def\epsfllx{#1}\ifx\epsfllx\epsfempty \epsfgrab #2 #3 #4 #5 .\\\else
   \global\def\epsflly{#2} \global\def\epsfurx{#3}\global\def\epsfury{#4}\fi}
\def\epsfsize#1#2{\epsfxsize} 
%
\setbox99=\hbox{\phantom{
\ref{\AfL} AffleckLudwig
\ref{\ARS} Alekseev, A.\ Recknagel, V.\ Schomerus
\ref{\ARStwo} fuzzy spheres 
\ref{\ABMNV} Alv-Gaume, Bost, 
\ref{\BirFS} Birke,Fuchs,Schwei
\ref{\BDLR} Brunner et al. 
\ref{\CLNY} C.G.\ Callan, C.\ Lovelace, C.R.\ Nappi, S.A.\ Yost,
\ref{\Carold} Cardy, 1984 and 1986 NPB works 
\ref{\Carfus} Cardy Fusionrules-paper
\ref{\CarLew} Cardy Lewellen 
\ref{\CIMM} Chen et al.: p=p' and B-field
\ref{\CorFai} Corrigan,Fairlie: off-shell states
\ref{\DiaRoem} Diac, Roemelsberger
\ref{\Dou}  Douglas
\ref{\DH} Douglas, Hull
\ref{\FFFSone} WZW branes 
\ref{\FM} Froehlich, Marchetti
\ref{\FSS} twining 
\ref{\FS} Fuchs, Schweigert : branes
\ref{\FSorbi} Fuchs, Schweigert, orbifold analysis
\ref{\FSgt} Fuchs, Schweigert: general theory
\ref{\FSms} Fuchs, Schweigert: more structures
\ref{\GNS} Gava, Narain, Sarmadi 
\ref{\GuS} Gutperle, Satoh
\ref{\Hash} hashimoto 
\ref{\Ish} Ishibashi
\ref{\Mal} Maldacena
\ref{\Na} Nahm quasi-rat 
\ref{\OOY} Ooguri, Oz, Yin
\ref{\Pol} Polchinski
\ref{\Rai} Raina 
\ref{\RSone} Recknagel, Schomerus, Gepner
\ref{\RStwo} RS moduli
\ref{\Sag} Sagnotti 
\ref{\Vol} Volker
\ref{\SW}
\ref{\Sen} Sen  
\ref{\Sta} Stanciu
\ref{\SV} Strominger, Vafa 
\ref{\Wibou} Witten, bound states 
\ref{\WiK}  Witten, K-theory
\ref{\Zamo} Zamolodchikov 
}}
%
{\nopagenumbers
\line{ETH-TH/99-29 \hfill HUTMP-99/B400}
\line{AEI-1999-39 \hfill  DESY 99-179} 
\leftline{hep-th/9912079}
\bn\bn\bn\bn
\centerline{{ \gros  {}Fundamental strings in Dp-Dq brane systems}} 
\bn\bn\bn 
\centerline{{\kap J\"urg Fr\"ohlich}}
\bn
\centerline{Institut f\"ur Theoretische Physik, ETH-H\"onggerberg,} 
\centerline{CH-8093 Z\"urich, Switzerland}
\bn\mn
\centerline{{\kap Olivier Grandjean}}
\bn
\centerline{Department of Mathematics, Harvard University, } 
\centerline{Cambridge, MA 02138, USA}
\bn\mn 
\centerline{{\kap Andreas Recknagel}}
\bn
\centerline{Max-Planck-Institut f\"ur Gravitationsphysik}
\centerline{Albert-Einstein-Institut}
\centerline{ Am M\"uhlenberg 1, D-14476 Golm, Germany}
\bn\mn 
\centerline{{\kap Volker Schomerus}}
\bn
\centerline{II. Institut f\"ur Theoretische Physik, Universit\"at Hamburg, } 
\centerline{Luruper Chaussee 149, D--22761 Hamburg, Germany }
\bn\bn\vfill
\bn\bn\bn
\centerline{{\bf Abstract}}
\bn
\narrower
We study conformal field theory correlation functions relevant 
for string diagrams with open strings that stretch between 
several parallel branes of different dimensions. 
In the framework of conformal field theory, 
they involve boundary condition changing twist fields which 
intertwine between Neumann and Dirichlet conditions. 
A Knizhnik-Zamolodchikov-like differential equation for correlators
of such boundary twist fields and ordinary string vertex operators 
is derived, and explicit integral formulas for its solutions are 
provided. 
\bn
\bn\bn
\vfill\vfill\vfill
\bn\bn
\quad e-mail addresses: {\tt juerg@itp.phys.ethz.ch, \ grandj@math.harvard.edu
\hfill \sn \quad\phantom{\rm e-mail addresses:}\ anderl@aei-potsdam.mpg.de, 
 \ vschomer@x4u.desy.de}   
\eject
\vfill $$\phantom{Leerseite}$$\vfill
\eject} 
\pageno=1  
\section{{1.\ \ Introduction}}%
D-branes \q{\Pol} have become the most important ingredient of 
the new picture of string theory that has emerged in recent years.  
They have shaped a new understanding of non-perturbative effects 
in string theory and of low-energy effective theories 
associated with string theories. In the latter context, systems 
of many branes are of particular importance, since they provide a 
natural way to include non-abelian gauge theories into string 
theory \q{\Wibou}. Systems of several branes of different dimensions, 
most notably of D1- and D5-branes, play a major role in proposals 
of how to derive the Bekenstein-Hawking entropy of black holes 
from string theory \q{\SV,\Mal}. More recently, stacks of branes and 
anti-branes have been reconsidered 
in connection with a $K$-theoretic classification of branes \q{\WiK}, 
based on results concerning tachyon condensation in \q{\Sen}.  
\hbn
Some qualitative features of such systems can be uncovered within 
a target space approach. But e.g.\ the process of brane-antibrane 
annihilation in the last-mentioned application, or the properties of 
near-extremal 
black holes involve the analysis of non-BPS states for which 
the world-sheet approach is better suited, since it does not 
critically depend on supersymmetry. Computation of CFT correlation 
functions is indispensable if one wants to deal with 
problems like Hawking radiation off D1-D5-systems, or for a 
clean discussion of bound state formation \q{\GNS}. 
\sn
The world-sheet formulation of string sectors with branes is 
well known in string theory, although mainly in connection 
with flat targets. The general setup involves boundary conformal 
field theory as introduced and developed by Cardy \q{\Carold,\Carfus,\CarLew}
and first exploited for string theory by Sagnotti \q{\Sag}. 
CFT on surfaces with boundaries exhibits a very interesting internal 
structure and finds interesting applications beyond string theory: 
For many s-wave dominated scattering processes, the universal behaviour 
is described by a boundary CFT in two dimensions, irrespective of 
the dimensionality of the original system. The most famous problem 
that could be tackled with boundary CFT methods is the Kondo effect in 
condensed matter physics \q{\AfL}. 
\hbn
In string theory, methods of boundary CFT are not only valuable 
in the study of situations without the BPS-property, but also 
to uncover non-classical features like unexpected moduli \q{\Sen,\RStwo} 
and non-commutative geometry; see e.g.\ \q{\DH,\Vol,\SW,\ARStwo,\FFFSone} 
and references therein. Moreover, they allow one to analyze 
D-branes in non-geometric string compactifications such as Gepner models 
\q{\OOY,\Sta,\RSone,\GuS,\BDLR,\DiaRoem}. 
\sn
In this paper, we ask how to compute CFT correlators describing 
string amplitudes of arbitrary closed and open string 
vertex operators in the presence of  multiple flat branes in $\R^D$. 
The open strings involved stretch between two {\sl or more} 
different branes, which may have {\sl different 
dimensions}. It appears that no systematic method for the computation 
of those string diagrams, which contribute to scattering processes 
in higher orders of the string coupling constant, is available in 
the literature. See, however, \q{\Hash} for some sample computations of 
scattering amplitudes in the presence of a pair of branes.
\sn
The world-sheet description requires surfaces with several boundary 
components. We restrict our attention to diagrams without internal 
closed string loops so that we can map the world-sheet to the disk 
or to the upper half-plane, but with different boundary conditions 
assigned to consecutive intervals on the boundary; see Figure 1. We focus 
on parallel branes here; thus we can reduce our analysis to a one-dimensional 
target. Results for $\R^D$ with $D>1$ follow by taking tensor products. 
The boundary state for a $p$-brane involves $p+1$ Neumann and $D-p-1$ 
Dirichlet boundary states of single free bosons. 
\hbn
The interesting transitions between boundary conditions in a one-dimensional 
target are those from Neumann to Dirichlet or vice versa. They are mediated 
by boundary fields of a special type, namely boundary condition 
changing {\sl twist fields}. 
\bn\mn
\vbox{
\hbox{\hskip-1.1cm\epsfbox{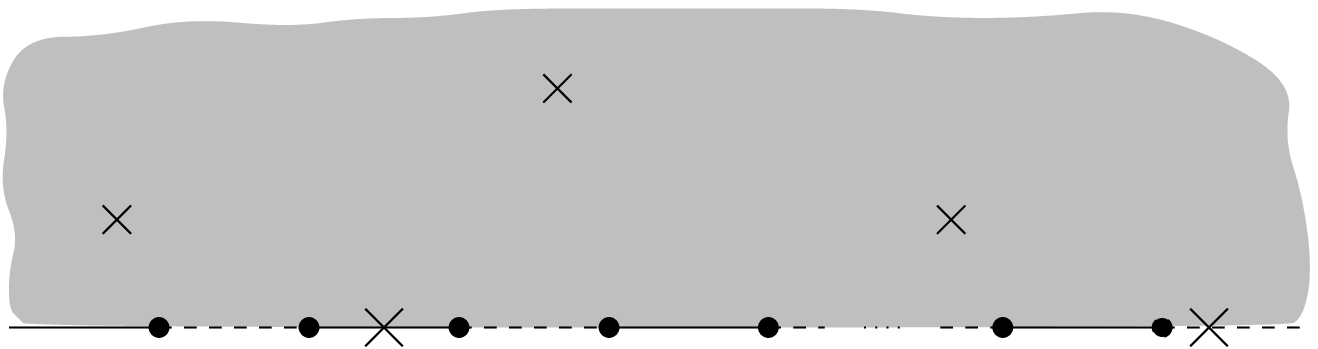}}
\sn\mn\mn
\narrower
{\bf Figure 1:} The upper half-plane with a sequence of Neumann (solid 
intervals) and Dirichlet (dashed intervals) boundary conditions along 
the real line. The dots between 
Neumann and Dirichlet intervals mark insertions of boundary condition 
changing twist fields, while crosses on the boundary or in the interior 
refer to insertions of ordinary open resp.\ closed string vertex operators.
Such a world-sheet diagram can be understood as Hawking radiation (closed 
string states) from a system of branes (multiple changes of boundary 
conditions) with simultaneous inner excitations (open string states).}
\bn\mn
Conformal boundary conditions which preserve the chiral algebra
$\cW$ of the theory are parametrized by certain automorphisms 
$\Omega$ of $\cW$,  together with the amplitudes of one-point 
functions \q{\RSone}. If $\cW$ is the U(1) current algebra, $\Omega$ 
can act as $\,\pm\,{\rm id}$ on the currents, and the 
1-point functions determine the location of a brane. 
If the boundary condition is constant along the 
boundary, arbitrary $n$-point functions can be expressed in terms
of the usual conformal blocks of $\cW$; see e.g.\ 
\q{\Carold,\RSone,\FS}. 
\hbn
If the gluing conditions described by an automorphism $\Omega$ of
$\cW$ remain constant along the boundary the computation of correlation 
functions is not, in principle, a difficult problem. 
Otherwise, the simple Ward identities for the symmetry 
algebra $\cW$ are broken, and one has to find new methods to 
construct the ``twisted chiral blocks'' involving a new type of 
boundary condition changing operators which correspond to 
twisted rather than ordinary representations of $\cW$. It is 
the aim of this article to develop a convenient formalism for 
computing such correlation functions in the case of a flat target space. 
\bn
The plan of this paper is as follows: In the next section, 
we look at correlation functions which contain just one  
insertion of a boundary twist field. We shall provide a 
complete operator construction of the boundary CFT, from 
which one can derive correlators with 
an arbitrary number of closed string vertex operators
inserted in the bulk. When there are more than two twist 
fields on the boundary, such techniques are no longer 
available. Our strategy is then to derive Ward identities 
for the correlation functions. They will lead us to 
Knizhnik-Zamolodchikov-like differential equations which 
describe the effect of moving insertion points for bulk 
and boundary fields in terms of a flat connection. We  
explain this idea in Section 3 and exploit it 
in the fourth section to give explicit integral formulas 
for the correlators. While some of the technical steps in 
setting up the Knizhnik-Zamolodchikov equations are rather 
involved, parts of our final results can be related to 
electrostatics. Section 5 comments on possible 
generalizations and applications.       
\bn
\section{2.\ \  Operator formalism for a single twist field insertion}%
As a simple example, we consider open strings propagating freely 
in the target $\R$, with Dirichlet boundary conditions imposed at one end 
of the string and Neumann boundary conditions at the other. 
We thus have to deal with a free bosonic field $X(t,\sigma)$
defined for space variables $\s \in [ 0, \pi ]$ and subject to 
$$
\partial_t X(t,0) = 0  \ \ \ \ \ \ , \quad \ \ \ \ 
\partial_{\sigma} X(t, \pi) = 0 
$$ 
for all $t\in\R$. Mapping the strip to the upper half-plane $\hh$ by 
$z= \exp(t+i\sigma)$, $X(z,\bar z)$ satisfies Dirichlet 
boundary conditions for $z\in\R_{>0}$ and Neumann boundary 
conditions for $z\in\R_{<0}\,$: 
$$ X(z,\bar z) \ = \ \der_0 \ \ \mbox{ for } \ \ z = \bar z  > 0 
    \ \ \ \ , \ \ \ \ 
  (\partial - \bar \partial)\,  X(z,\bar z) \ = \ 0 \ \ \mbox{ for } 
  \ \ z = \bar z < 0 \ \ . 
\eq\bdconds$$
Our task is to compute correlation functions in this bosonic 
theory which involves two insertions of twist fields on the 
boundary (see \q{\CorFai} for an early treatment of that problem). Conformal symmetry allows us to place these boundary 
condition changing operators at $x_1 = 0$ and $x_2 = \infty$.
\hbn 
We propose to construct operators $X(z, \bar z)$ satisfying \bdconds, 
as well as open and closed string vertex operators, and then 
to derive differential equations on the correlation functions from 
the algebraic properties of these operators and from the symmetries 
of the theory. First, we have to 
determine the space our fields are to act on. 

\subsection{2.1\ \ The spectrum of boundary twist fields}  
The space in question is spanned by excited states of open 
strings stretching between a Neumann and a Dirichlet boundary 
condition -- hence the name ``boundary condition changing 
operators'' for the boundary fields uniquely associated to these 
states. There is a relatively simple technique to determine 
the spectrum of boundary condition changing operators 
that intertwine between two constant conformal boundary 
conditions $B_1 = (\Omega_1,\alpha_1)$ and $B_2 = 
(\Omega_2,\alpha_2)$ in some boundary CFT. 
The state space of this boundary theory 
is denoted by $\cH_{12}$. Because of the state-field 
correspondence, the spectrum of boundary condition changing operators 
is described through the partition function of the boundary 
theory, 
$$ Z_{12} (q) \ = \ {\rm Tr}_{\cH_{12}}\, q^{H\uh}\ \ \ \  
   \mbox{ where } \ \ \ \ H\uh \, = \, L_0 - { c \over 24} \ \ .$$
By an interchange of space and time coordinates (``world-sheet 
duality''), the open string 1-loop diagram underlying $Z_{12}$ may 
be viewed as a closed string tree diagram, i.e.\ 
$$ Z_{12} (q) \ = \ \langle B_1|\, \tilde q^{\frac12 H\up} 
   \, |B_2\rangle\ \ \ \ \mbox{ where } \ \ \ \ H\up \, = \, 
    L_0\up + \bL_0{}\up - \frac{c}{12} \ \ , $$
where $\tilde q= \exp(- 2 \pi i / \tau)$ is related to the 
variable $q = \exp(2\pi i \tau)$ as usual. The closed strings propagate 
between the {\it boundary states} $|B_i\rangle 
= |\alpha_i\rangle_{\Omega_i}$ associated 
with the boundary conditions $B_i$. They allow to transfer 
boundary conditions from the upper half-plane (where the Hamiltonian 
is $H\uh$) into a CFT on the full plane (with Hamiltonian  $H\up$), 
see \q{\Ish,\Carfus}. 
\sn
The boundary states implementing Dirichlet and Neumann conditions along 
the whole boundary are of course well known, see e.g.\ \q{\CLNY}. 
Let $\ja_n\up,\ \jba{}_n\up$ 
be two commuting sets of oscillator modes (in the plane CFT) 
with standard commutation relations. The ground states $|k\rangle$ of 
their Fock spaces are labeled by the momentum $k\in \R$. Neumann 
and Dirichlet boundary states are given by 
$$\eqalign{
&|\,N\,\rangle  = {1\over \sqrt{2}}\,
   \exp\bigl\{\, - \sum_{n\geq 1} {1\over n}\, 
     \ja_{-n}\up\,\jba{}_{-n}\up\,\bigr\}\; |0\rangle \ , 
\cr
&|\,D(\der_0)\,\rangle  = \int\!dk\; e^{ik \der_0} \,
      \exp\bigl\{\,  \sum_{n\geq 1} {1\over n}\, 
     \ja_{-n}\up\,\jba{}_{-n}\up\,\bigr\}\; |k\rangle \ , 
\cr}
\eq\bdstates$$
where $\der_0 \in \R$, as in \bdconds, denotes the location of the 
``D-brane'', i.e.\ $\hat x\,|D(\der_0)\rangle = \der_0\,|D(\der_0)\rangle$
for the center of mass coordinate $\hat x$. 
\hbn
If we build the boundary state $|p\rangle$ for a $p$-brane in $\R^D$ 
as a tensor product of $|N\rangle$ ($p+1$ times) and $|D(\der^i_0)\rangle$ 
($i=p+2,\ldots,D$) from eq.\ \bdstates, the partition function $Z_{p\!p}(q)$
counts boundary fields that do not change the boundary condition. They 
describe excitations of open strings attached to the $p$-brane. 
These open string vertices have the form $\varepsilon_{\mu_1\ldots\mu_n}
\partial X^{\mu_1}\cdots\partial X^{\mu_n} e^{ikX}$ with certain 
Lorentz tensors $\varepsilon_{\mu_1\ldots\mu_n}$ and with momentum $k$ 
parallel to the Neumann directions. The case $n=1$ (where the polarization 
$\varepsilon_\mu$ is transversal) contains the massless modes: gauge 
fields living on the brane world-volume. 
\mn
The partition function of the theory with Neumann 
boundary conditions on one side and Dirichlet on the other 
follows from the boundary states as explained above, 
$$\eqalignno{
Z_{N\!D}(q) &= \  {\rm Tr}_{{\cal H}_{N\!D}}\; q^{H_{N\!D}} = 
  \langle \,N \,|\, \tilde q^{L_0 - {c\over24}} \, |\,D(\der_0)\,\rangle
&\bcr 
&= \ {1\over \sqrt{2}}    \langle 0 | \, 
e^{- \sum_{n=1}^\infty {1\over n}\,\ja_{n}\,\jba_{n}} \, 
\tilde q^{L_0 - {c\over24}} \,
 e^{\sum_{m=1}^\infty {1\over m}\,
        \ja_{-m}\, \jba_{-m}}\ |0\rangle\ \ .
&\mkg\zndtilexp\cr }$$
Orthonormality of the Fock ground states implies that only the 
contribution from the vacuum sector survives in the second line.
In particular, $Z_{N\!D}$ is independent of the 
parameter $\der_0$. Computation of the vacuum expectation value 
above is straightforward. The result can be written as 
$$
Z_{N\!D}(q) = q^{{1\over48}} \prod_{n=1}^{\infty} (1-q^{n-{1\over2}})^{-1}
= {1\over \eta(q)}\; \sum_{n=1}^{\infty} q^{{1\over4}(n-\oh)^2}\ .
\eq\ZND$$
Our main conclusion concerns the conformal weights of the boundary 
fields that can induce a transition between Dirichlet and Neumann
type boundary conditions. The lowest weight that appears is 
$h={1\over 16}$. Above this
value, the spectrum of conformal weights has half-integer 
spacings. 
The boundary condition changing operator  with conformal weight 
$h = {1\over 16}$ corresponding to the lowest-energy state 
$|\s\rangle$ in the whole sector $\cH_{N\!D}$ will be called  $\s(x)$.
\hbn
We will also refer to $\sigma(x)$ as a ``twist field'' since the 
sum of irreducible Virasoro characters in \ZND\ can alternatively 
be regarded as the character of a twisted U(1) representation. The 
absence of a vacuum state and the half-integer energy grading 
are symptoms for the fact that the jump from Neumann to Dirichlet 
destroys the simple U(1) Ward identities that are present in a boundary 
CFT with constant Neumann or Dirichlet condition all along the 
boundary. See \q{\FSgt} for general results about twist fields and 
partition functions in boundary conformal field theory.
\hbn
It will be our main concern in the following to find ``substitutes'' 
for the broken Ward identities, namely twisted Knizhnik-Zamolodchikov 
equations. 
\bn 
\subsection{2.2\ \  Construction of the basic fields} 
In order to construct a field $X(z,\bar z)$ obeying the
boundary conditions \bdconds\ 
on $\cH = \cH_{N\!D}$, we introduce a set of oscillator 
modes $a_r$ labelled by half-integers $r \in \Z +\frac12$. 
(We drop the superscript ${}^{(H)}$ for operators of the 
upper half-plane theory.) 
They are supposed to obey the relations
$$   [\, a_r\, , \, a_s\, ]\ = \ r \, \d_{r,-s} \ \ \ \ , \ \ \ 
     a_r^* \ = \ a_{-r} \ \ . $$
Creation operators $a_r,\ r < 0,$ generate the Fock space 
$\cH$ out of the ground state $\!\vac$, which is annihilated by 
modes $a_r$ with index $r >0$.  All the fields 
we shall consider act on this state space $\H$. It is 
simple to verify that the decomposition $X(z,\bar z) =  X(z) - \bX(\bz)$
of the bosonic field yields the desired properties if 
$$X(z) \ = \ \der_0 \, + \, i \sum_{r \in \mZh} \ 
  \frac{a_r}{r}\ 
   z^{-r}  \ \  , \ \ \
   \bX(\bz) \ = \ i \sum_{r \in \mZh} \ \frac{a_r}{r} \ \bz{}^{-r}\ \ .
$$
To make the square root well defined, we have to introduce a branch cut 
in the plane, which extends from  $x=0$ to $-\infty$. 
Once the bosonic field is known, we obtain chiral currents as 
$$ J(z) \ := \ i \, \partial X(z,\bz) = \ \sum_{r \in \mZh} 
   \, a_r \ z^{-r-1} \ \ \ , \ \ \  
   \bJ(\bz) \ := \ i\, \bar \partial X(z,\bar z) = \ - 
   \sum_{r \in \mZh}\, a_r \ \bz^{-r-1} \ \ . $$
Finally, the components $T(z)$ and $\bT(\bz)$ of the stress 
energy tensor are given by 
$$ T(z)\  = \ \lim_{w \rightarrow z}\  \frac12\left( 
        J(w) J(z) - \frac{1}{(w-z)^2}\right) \ \ , 
$$ 
and likewise for $\bT(\bz)$. Since $T$ and $\bT$ are 
quadratic in $J$ and $\bJ$, they satisfy the usual 
boundary condition $T(z) = \bT(\bz)$ all along the 
real line $\Im z = 0 $.  By the usual arguments \q{\Carold} 
this implies that the modes
$$ L_n\ := \ \int_{C_{+}} \frac{dz}{2\pi i}\; z^{n+1} \,T(z) +
         \ \int_{C_{-}} \frac{d\bz}{2\pi i}\; \bz^{n+1}\, \bT(\bz) 
  \ \ \ 
$$
obey commutation relations of the Virasoro algebra with central charge 
$c = 1$. Here $C_{+}-C_{-}$ is a closed oriented contour surrounding 
the origin, with $C_{+}$ contained in the upper half-plane and $C_{-}$ 
contained in the lower half-plane. The commutation relation 
between $L_n$ and $a_r$ is easily checked to be of the form 
$$ [\ L_n\, ,\, a_r\ ] \ = \ -r\  a_{n+r} \ \ . $$ 
It is convenient to introduce two generating fields 
$\cT$ and $\cJ$ by the formal sums
$$ \cT(\o) \ = \ \sum_{n \in \mZ}\, L_n \ \o^{-n-2} \ \ , \ \ 
   \cJ(\o) \ = \ \sum_{r \in \mZh}\,  a_r \ \o^{-r-1} \ \ .$$
One may think of $\cT$ as being defined on the entire complex plane 
with $\cT(\o) = T(\o)$ in the upper half-plane, $\Im \o>0$, 
and $\cT(\o) = \bT(\o)$, for all $\o$ with $\Im \o < 0$. The 
generating field $\cJ$ naturally lives on the two-fold branched 
cover of the complex plane defined by $\ome^2 = \o$. By introducing 
the branch cut from $x=0$ to $-\infty$ we have specified a 
coordinate patch on this surface with the local coordinate  
being denoted by $\o$. In this chart, $\cJ(\o) = J(\o)$ 
for $\Im \o>0$ and $\cJ(\o) = - \bJ(\o)$ for $\Im \o < 0$. 
\mn
The commutation relations for the modes $L_n, a_r$ with the 
bosonic field $X(z,\bz)$ can be expressed in terms of 
$\cT$ and $\cJ$ as follows 
$$\eqalign{ 
&\phantom{xxxxxxx}\vphantom{\sum_X}[\ \cT(\o)\, , \, X(z,\bz)\ ]   
  =   \p X(z,\bz)\ \d(z -\o) + \bp X(z,\bz) \ \d(\bz - \o) \ \ , \nn 
\cr 
&\phantom{xxxxxxx}\vphantom{\sum_X}[\ \cJ(\o)\, , \, X(z,\bz)\ ]   
=    i\,  \d(z-\o)  \ +  \ i\, \d(\bz- \o) \ \ , \nn
\cr
\noalign{\leftline{\rm where}} 
&\phantom{xxxxxxx\sum_X^X}\d(z-\o)\  :=  \  \frac{1}{z}\,  \sum_{n\in\mZ}  
    \, \left( \frac{z}{\o}\right)^n \ = \ \frac{1}{z}\, \sum_{r\in \mZh} 
     \, \left( \frac{z}{\o}\right)^r\ \ . \nn  
\cr}$$
We state two simple consequences of these formulas that 
shall be important below. We split $\cJ$ and $\cT$ 
into two parts $\cJ(\o) = \cJ_> (\o) + \cJ_< (\o)$ and 
$\cT(\o) = \cT_>(\o) + \cT_< (\o)$ such that 
$$   \cJ_>(\o ) \ := \ \sum_{r \geq 1/2} \ \ a_r \ \o^{-r-1} 
       \ \ \ , \ \ \ 
  \cT_>(\o ) \ := \ \sum_{n \geq -1}\ L_n \ \o^{-n-2} \ \ . $$ 
In the next subsection, we will use the commutation relations between 
the singular parts $\cT_>, \cJ_>$ of the generating fields $\cT,\cJ$ 
and the bosonic field $X(z,\bz)$: 
$$ 
[\, \cJ_>(\o)\, ,\, X(z)\, ]  =     
 - \left( \frac{z}{\o}\right)^{1/2} \frac{i}{\o-z} \ ,
\quad\quad
[\, \cT_>(\o)\, ,\, X(z)\,]  =  \frac{1}{\o - z}\  \p_z X(z) \ . 
\eq\TX$$ 
Moreover, we will need the following lemma, a proof of which is 
given in Appendix A. 
\bn
\noindent
{\bf Lemma 1:}\quad \it  
One may rewrite the generating field $\cT(\o)$ in terms of the 
objects $\cJ_>(\o)$ and $\cJ_<(\o)$, namely 
\be 
\cT(\o) \ = \ \frac{1}{2} \bigl(\, \cJ_< (\o) \cJ(\o) 
  + \cJ (\o) \cJ_> (\o)\, \bigr) + 
   \frac{1}{16} \frac{1}{\o^2} \ \ . \labela\Sug 
\ee
\rm  
\bn 

\subsection{2.3\ \  Bulk and boundary primary fields}  
Our next aim is to construct primary bulk and boundary 
fields. Here, the latter term refers to open string 
vertex operators which can be inserted on $\R_{<0}$ or $\R_{>0}$
without changing the boundary condition. They are in one-to-one 
correspondence to states in $\cH_{D\!D}$ or $\cH_{N\!N}$, {\sl not} to 
states in $\cH_{N\!D}$. We will see that bulk fields can be 
regarded as products of such boundary fields $\Psi_\rg(z)$. 
Therefore we discuss these ``chiral fields'' first -- 
admitting arbitrary complex insertion points, not just 
$z \in \partial \hh$. The fields $\Psi_\rg(z)$ are labeled 
by a real parameter $\rg$ and enjoy the properties 
\ba{ 
[\, \cT(\o)\, ,\, \Psi_\rg(z)\, ] & \  = \   
 \p_z \Psi_\rg(z) \ \d(x-\o) + h \, \Psi_\rg(z) \ \p_z
 \d(z-\o)\ \ ,\phantom{\sum_i} \nn \cr
[\, \cJ(\o) \, , \, \Psi_\rg(z)\, ] & \ = \   \rg\, \Psi_\rg(z)\ 
  \d(z-\o) \ \ . \nn
\cr }\ea 
We have used the definition of the formal $\d$-function specified above and 
$h = \frac12 \rg^2$. For the commutators of the $\cT_>(\o),\cJ_>(\o)$ 
with the fields $\Psi_\rg(x)$, this implies 
\ba{ 
[\, \cT_>(\o)\, ,\, \Psi_\rg(z)\, ] & \ =\    
 \frac{1}{\o - z}\  \p_z \Psi_\rg(z)  + 
      \frac{h}{(\o -z)^2} \ \Psi_\rg(z)\ \ , \labelb\TP \mkgho\TPho \cr
[\, \cJ_>(\o) \, , \, \Psi_\rg(z)\, ] & \ = \   
\left( \frac{z}{\o}\right)^{1/2} \frac{\rg}{\o-z} \ \Psi_\rg(z) 
 \ \ . \labelb\JP\mkgvo\JPvo 
\cr }\ea 

\noindent
{\bf Lemma 2:}\quad \it The unique solution (up to normalization)
$\Psi_\rg(z)$ to the requirements \TPho,\JPvo\  
is given by 
$$ \Psi_\rg(z) \  =  \ \no z^{-h} \, e^{i \rg X_<(z)} \ e^{i \rg X_>(z)} 
    \ \ \ \ \mbox{ where } \  \ \ \   X_> (z) 
    \ = \ i \sum_{r>0} \, \frac{a_r}{r}z^{-r}
$$
and $X_<(z) = X(z) - X_>(z)$. Note that $\Psi_\rg(z)$ 
is normal-ordered, i.e., the annihilators $a_r, r>0,$ appear 
to the right of the creation operators.  
\rm
\bn 
A proof can be found in Appendix A. 

\noindent
Our next aim is to describe 
the U(1)-primary bulk fields $\phi_\rg(z,\bz)$. By definition, they obey 
the following commutation relations with  respect to $\cJ_>$ and 
$\cT_>$,
\ba{ 
[\, \cT_>(\o)\, ,\, \phi_\rg(z,\bz)\, ] & =   
 \frac{1}{\o - z}\  \p \phi_\rg(z,\bz)  + 
      \frac{h}{(\o -z)^2} \ \phi_\rg(z,\bz) \labelb\Tp \cr   
      &   \hspace{1.5cm} + \frac{1}{\o - \bz}\  \bp \phi_\rg(z,\bz)  + 
      \frac{h}{(\o -\bz)^2} \ \phi_\rg(z,\bz) \ , \nn \cr
[\, \cJ_>(\o) \, , \, \phi_\rg(z,\bz)\, ] & =   
\left( \frac{z}{\o}\right)^{1/2} \frac{\rg}{\o-z} \ \phi_\rg(z,\bz) - 
\left( \frac{\bz}{\o}\right)^{1/2} \frac{\rg}{\o-\bz} \ \phi_\rg(z,\bz) 
\ . \labelb\Jp 
\cr }\ea  
Note that each term from eqs.\ \TPho,\JPvo\ appears a second time with 
$z$ being replaced by the variable $\bz$. One can easily work out 
commutation relations between the full generating elements 
$\cT(\o),\cJ(\o)$ and the bulk primary fields $\phi_\rg(z,\bz)$.
It is obvious from our discussion of boundary fields that bulk fields 
$\phi_\rg(z,\bz)$ can be written as products of chiral vertex operators,  
$$ \phi_\rg(z,\bz)\ = \ \Psi_\rg(z)\, \Psi_{-\rg}(\bz) \ \ . $$ 
\mn
The formulas we have reviewed here would enable us to perform a direct 
computation of arbitrary correlations functions $G(\bfz)$ for bulk-fields 
$\phi_\rg(z,\bz)$ with two twist fields $\s$ inserted at $x=0$ and 
$x = \infty$, 
$$ G(\bfz) \ := \ \langle \s | \, \phi_1(z_1, \bz_1) \dots 
   \phi_n(z_n,\bz_n)|\s\rangle \ \ \ \mbox{ where } \ \ \   
   \phi_\nu(z_\nu,\bz_\nu) \ = \ \phi_{\rg_\nu} (z_\nu, 
    \bz_\nu) \ \ . $$
The calculations would proceed by moving all annihilation operators
to the right until they act on the ground state $\!\vac$. The same 
techniques would apply if there are extra boundary fields $\Psi_\rg(x)$ 
inserted in addition to the bulk fields $\phi_\rg(z,\bz)$. Since we will 
develop another, more generally applicable approach to the computation 
of correlation functions below, we do not enter details here. 
\bn
Before we conclude this subsection, we would like to derive 
{\sl bulk-boundary operator product expansions} which allow to expand 
our bulk fields $\phi_\rg(z,\bz)$ in terms of boundary operators 
\q{\CarLew}. The essential idea is that, in the presence of a boundary, 
bulk fields split into products of chiral vertex operators inserted at 
points which are obtained {}from each other by reflection at the real 
axis and with opposite charges (``method of image charges'').
In order to obtain concrete formulas, we first rewrite the bulk 
fields $\phi_\rg(z,\bz) = \Psi_\rg(z) \Psi_{-\rg}(\bz)$ in terms of 
$X_>(z,\bz) = X_>(z) - \bX_>(\bz)$ and $X_< (z,\bz) = X_<(z)-\bX_<(\bz)$, 
$$
\phi_\rg(z,\bz) = \ \noq (z\bz)^{-h} 
   \left(\frac{\sqrt{z} + \sqrt{\bz}}
   {\sqrt{z}-\sqrt{\bz}}\right)^{2h}
   \ e^{i\rg  X_<(z,\bz)} \ e^{i\rg  X_>(z,\bz)} \ \ . 
\labela\phie
$$
We have used the expression in Lemma 2 and then normal-ordered 
the right hand side with the help of the BCH formula, which leads 
to the additional $\sqrt{z}$- and$\sqrt{\bz}$-dependent factor. 
\bn
\noindent
{\bf Lemma 3}\ \it {\rm (bulk-boundary OPE)}{\bf :} \quad For 
arguments $z = x+ i y$ close to the boundary, i.e., $y>0$ small, 
the operators $\phi_\rg (z,\bz)$ can be expanded in a series 
involving boundary primary fields, with leading asymptotics  
\ba{ 
  \phi_\rg (z,\bz) & \sim   \frac{e^{i\rg \der_0}}{y^{2h}}
    \ \ {\bf 1} \ \ \quad\quad\quad   
  \mbox{ for } \ \ \ x > 0 \ \ , \nn \bcr
   \phi_\rg (z,\bz) & \sim   y^{2h} \ \Psi_{2\rg }(x) \ \quad\quad 
  \mbox{ for } \ \ \ x < 0 \ \ , \nn 
\cr }\ea
where ${\bf 1}$ is the identity field. 
\rm 
\bn    
\noindent
{\sc Proof:} 
Let us begin with the case $x>0$ in which $ \sqrt{z} - 
\sqrt{\bz} \rightarrow 0 $ as $y$ becomes very small.
\ba{ 
\phi_\rg (z,\bz) & =   \noq (z\bz)^{-h} \left( 
\frac{ z+\bz + 2\sqrt{z\bz}}{z-\bz} \right)^{2h} 
  e^{i\rg X_<(z,\bz)}e^{i\rg X_>(z,\bz)} \nn \cr 
 & \sim   \noq x^{-2h}\  \left(\frac{2x}{iy}\right)^{2h} \ 
     {\bf 1} \ = \ \frac{1}{y^{2h}} \ {\bf 1 } \ \ . \nn
\cr }\ea
We have also used that $X_<(x,x) = \der_0$ and $X_>(x,x) = 0$ for 
$x > 0$, which is  a direct consequence of the Dirichlet boundary 
condition. 
\sn
If $x<0$ and $y$ tends to zero, the sum $\sqrt{z}+\sqrt{\bz}$ 
vanishes and we can estimate the behaviour of $\phi_\rg (z,\bz)$ 
according to    
$$\eqalignno{ 
\phi_\rg (z,\bz) &=   \noq (z\bz)^{-h} \left( 
\frac{z-\bz }{z+\bz -2\sqrt{z\bz}} \right)^{2h} 
  e^{i\rg X_<(z,\bz)}e^{i\rg X_>(z,\bz)} &\cr 
 & \sim   \noq x^{-2h}\  \left(\frac{iy}{2x}\right)^{2h} \ 
   e^{2i\rg  X_<(x)} e^{2i\rg  X_>(x)} \ = \ y^{2h} \ \Psi_{2\rg }(x) \ \ . 
&\qed\cr }$$
Observe that boundary condition changing operators themselves do not 
arise from the bulk-boundary OPE of bulk fields. Let us finally note that 
operator product expansions for bulk fields or for boundary fields can be 
worked out with the same techniques. Those for bulk fields 
$\phi_\rg (z,\bz)$ of course agree with the usual OPE of primary fields 
in the bulk.

\subsection{2.4\ \  Correlation functions and the Knizhnik-Zamolodchikov 
equation} This subsection contains the main result of this section, 
namely a derivation of the Knizhnik-Zamolodchikov equation for 
correlation functions of bulk and boundary primaries in the presence 
of a transition from Dirichlet to Neumann boundary conditions 
at the origin. Let us look more closely at correlation functions 
containing $n$ chiral fields $\Psi_\rg (z)$,   
$$ F(\bfz) \ := \ \langle \s|\, \Psi_1(z_1) \dots \Psi_n(z_n) 
  |\s\rangle \ \ \ \ \mbox{ where } \ \ \ \ \Psi_\nu (z_\nu) 
  \ := \  \Psi_{\rg _\nu}(z_\nu) \ \ . $$
Before we start, let us state two elementary formulas for the action 
of $\cJ_>(\o)$, $\cT_>(\o)$ on the ground state $\!\vac\,$:  
\ba{ 
\cT_>(\o)\  \vac & \ = \  \left( \frac{1}{\o^2} \, h_\sigma 
    + \frac{1}{\o}  \, L_{-1}  \,  \right) \, \vac \ \ , 
\labelb\Tv  
\cr 
\cJ_>(\o) \ \vac & \ = \ \sum_{r\geq 1/2}^{\phantom{k}} \, a_r  \vac 
   \; \o^{-r-1} \ = \ 0  \ \ ,     
\labelb\Jv
\cr }\ea    
where $h_\sigma$ is the conformal weight of the state $\!\vac$. 
We will recover the equation $h_\s = {1\over16}$ in a moment. 
The object $\cav\!$ dual to $\!\vac$ obeys the relations $\cav 
\cJ_<(\o) = 0 = \cav \cT_<(\o)$. 
\mn
A first differential equation is obtained by inserting the 
generating field $\cT(\o)$ into the correlation function: 
\ba{ 
\lefteqn{\cav \cT(\o) \Psi_1(z_1) \dots \Psi_n(z_n)\vac  \ = \ 
  \cav \cT_>(\o) \Psi_1(z_1) \dots \Psi_n(z_n)\vac}  \nn \bcr
& =  \ \left[\; \sum_{\nu=1}^n \left( \frac{1}{\o-z_\nu}\ \partial_\nu 
    \ + \ \frac{h_\nu}{(\o-z_\nu)^2} \,\right) + \frac{h_\sigma}{\o^2} 
   \ \right] \ F(\bfz)  + \frac{1}{\o} \ \cav \Psi_1(z_1) \dots 
    \Psi_n(z_n)\, L_{-1}  \vac\ \ . \nn
\cr }\ea 
Here, we have commuted $\cT_>(\o)$ through the fields $\Psi_\nu(z_\nu)$ 
until it acts on the ground state $\!\vac$ so that we can 
use formula \refe{\Tv}.
\hbn
Now we want to compute the same correlation function with the help of 
the affine Sugawara construction, i.e.\ by exploiting eq.\ \refe{\Sug}: 
\ba{ 
 \lefteqn{\cav \cT(\o) \Psi_1(z_1) \dots \Psi_n(z_n)\vac} 
  \nn \phantom{\sum_i} \cr 
  & \ =  \ \frac12 \, \cav \cJ_>(\o) \cJ_>(\o)  \Psi_1(z_1) \dots 
    \Psi_n(z_n)\vac + \frac{1}{16\o^2} \,\cav  \Psi_1(z_1) 
    \dots\Psi_n(z_n)\vac \nn \phantom{\sum_K^K} \cr
& \ = \   \frac12 \, \sum_{\nu=1}^n \left(\frac{z_\nu}{\o}\right)^{1/2} 
     \frac{\rg _\nu}{\o-z_\nu}\ \cav \cJ_>(\o) \Psi_1(z_1) \dots 
     \Psi_n(z_n)\vac +  \frac{1}{16\o^2} \, F(\bfz)  \nn 
      \phantom{\sum_K^K} \cr
& \ =  \  \left[ \frac12\,  \sum_{\nu,\mu} \frac{\sqrt{z_\nu z_\mu}}{\o} 
      \frac{\rg _\nu \rg _\mu}{(\o-z_\nu)(\o-z_\mu)}  + 
      \frac{1}{16 \o^2} \right] \ F(\bfz) \ \ . \nn   
\cr }\ea 
Comparison with our first formula for the insertion of $\cT(\o)$ 
yields $h_\sigma = \frac{1}{16}$. From the residue at $w=0$ we get 
$$  
  \cav \Psi_1(z_1) \dots 
              \Psi_n(z_n)\, L_{-1} \vac\ \ = \ 
   \frac12\ \sum_{\nu,\mu} \frac{\rg _\nu \rg _\mu}{\sqrt{z_\nu z_\mu}} 
    \ F(\bfz)  \ \ . 
$$ 
Finally, from the residue at $\o = z_\nu$ we obtain 
\be 
\p_{z_\nu} \, F(\bfz) \ = \  
\left[ - \frac{h_\nu}{z_\nu}\ +\ \sum_{\mu\neq \nu} 
   \sqrt{\frac{z_\mu}{z_\nu}}\; 
  \frac{\rg _\nu \rg _\mu}{z_\nu - z_\mu}   \right]\ F(\bfz)\ .  
\labela\simpleKZeq\ee
This is the {\sl Knizhnik-Zamolodchikov equation} we were after. 
Note that the terms in square brackets determine a flat connection,  
as in the ordinary Knizhnik-Zamolodchikov equation. 
\sn
We can solve \simpleKZeq\ by a simple coordinate transformation. 
In fact, if we introduce coordinates $u_\nu = \sqrt z_\nu$ and 
the function $F_u(u_1, \dots, u_n) = \prod_\nu u_\nu^{h_\nu} 
F(u^2_1,\dots,u^2_n)$, then the system \simpleKZeq\ of first order 
differential equations becomes         
$$
\partial_{u_\nu}\,F_u(u_1,\ldots, u_n) = 
\biggl( - {\rg _\nu^2 \over 2 u_\nu} \; +\;  
\sum_{{\mu=1 \atop  \mu\neq\nu }}^n 
\Bigl[ {\rg _\nu \rg _\mu \over u_\nu-u_\mu} 
        - {\rg _\nu \rg _\mu \over u_\nu +u_\mu }  \Bigr] 
\biggr) \,F_u(u_1,\ldots, u_n )\ .\phantom{xxx}
\eq\kzu$$
This equation is formally identical to the usual Knizhnik-%
Zamolodchikov equation with $2n$ fields of charges $\pm \rg _\nu$ 
inserted at the points $\pm u_\nu$.  The solution to \kzu\ is given by 
$$
F_u(u_1,\ldots, u_n) = \kappa\cdot \prod_{\nu=1}^n u_i^{-{\rg _\nu^2\over2}}\,
\prod_{1\leq \nu<\mu \leq n} \Bigl( {u_\nu-u_\mu \over u_\nu+u_\mu} 
 \Bigr)^{\rg _\nu \rg _\mu}\ . 
\eq\kzusol$$ 
The free parameter $\kappa$ can be determined from the boundary 
condition of the bosonic field $X$ on the positive real line, i.e.\ 
$\kappa = \kappa(\der_0)$ depends on the position $\der_0$ of the
D-brane.
\bn
\section{3.\ \ Twisted Knizhnik-Zamolodchikov equation for multiple 
\phantom{xn} transitions}%
In the following, we study $n$-point functions of a free 
bosonic field theory on the half-plane with several insertions 
of twist operators placed along the boundary. In the corresponding 
string diagrams, open strings stretch between three or more branes 
of various dimensions.  
\hbn
As long as only one DN-jump occurs, a simple Hilbert space formulation 
of the boundary CFT is available, and we can solve, in principle, for 
correlation functions by purely algebraic techniques, as indicated in 
the last section. In the presence of many boundary condition changing 
twist fields, it may be simpler to resort to OPE methods and to the theory 
of complex functions on higher genus Riemann surfaces, and this is the 
approach we pursue in the present section. We begin with a
very brief review of  relevant input from the theory of
hyperelliptic surfaces. We then discuss Ward identities in
the second subsection. The latter allow us to derive a system of
linear first order differential equations for the correlation
functions similar to the Knizhnik-Zamolodchikov equations.
Note that free bosons on higher genus surfaces without boundaries 
(i.e.\ higher loop diagrams of closed strings propagating in a 
flat target) have been studied in great detail in \q{\ABMNV,\Rai}. 

\subsection{3.1\ \  Hyperelliptic surfaces} 
Our aim is to investigate a scenario in which a bosonic field 
$X(z,\bar z)$ is defined on the upper half-plane with boundary
conditions switching between Dirichlet and Neumann at $2g+2$
points $x_i, i = 1, \dots, 2g+2,$  on the boundary. Without loss
of generality, we shall assume that $x_{2g+2} = \infty =: x_0$.
To be more precise, we impose Dirichlet boundary conditions in
the intervals $]x_i,x_{i+1}[$ for $i$ odd and Neumann boundary
conditions along the rest of the boundary, i.e.,
\ba{ 
X(z,\bz) \  & = \ \der^k_0 \ \ \ \mbox{\ for\ } \ \ \ z \, = \, \bz 
    \ \in\ D_k \ :=\  ]x_{2k-1},x_{2k}[ \ \ \ \ \cr
\noalign{\leftline{\rm and}} 
    \p_y X(z,\bz) \  & = \ 0 \ \ \ \mbox{\ for \ } \ \ \ z \, = \, \bz
   \ \in \ N_k \ :=\ ]x_{2k-2},x_{2k-1}[ \ \ \ \ . \cr}    
\ea
The variable $y$ is defined through $z = x+iy$, and $k = 1, \dots, g+1$.
In terms of the chiral currents $J(z) = i \, \partial X(z,\bar z)$ and 
$\bJ(\bar z) = i\, \bar \partial X(z,\bar z)$,  these conditions become 
$$ J(x) \ = \ - \bJ(x) \ \ \ \mbox{\ for\  } \ \ \ x \ \in \ D_k\ ,
   \ \ \ \mbox{\ and \ } \ \ \ \  
   J(x) \ = \ \bJ(x) \ \ \mbox{\ for\ }\ x \ \in \ N_k
   \ \ \ \ . 
$$
As in the previous section, it is convenient to work with a single field 
$\cJ$ that contains all information about the two chiral currents
$J$ and $\bJ$. Such a field necessarily lives on a two-fold branched 
cover of the complex $w$-plane, namely on the {\sl hyperelliptic 
surface}, $M$, of genus $g$ which is described by the equation 
$$  \ome^2 \ = \ P(\o) \ :=\  \prod_{i=1}^{2g+1} (\o - x_i) \ \ . $$
Introducing branch cuts along the intervals $N_k = [x_{2k-2},x_{2k-1}]\,$, 
we obtain a particular coordinate patch of this surface with local 
coordinate $w$. In this chart, $\cJ(w)$ satisfies $\cJ(w) = J(w)$ 
for $\Im w>0$ and $\cJ(w) = - \bJ (w)$ for $\Im w< 0$.  
The Virasoro field $T$ obeys the gluing condition $T(x) = \bT(x)$, 
all along the boundary, since it is quadratic in the currents. 
Consequently, the generating field $\cT(w)$ is defined on the 
complex $w$-plane and coincides with $T$ (resp.\ $\bT$) on the 
upper (resp.\ lower) half-plane. 
\mn
The coordinate $w$ on the complex plane lifts to a meromorphic
function of degree $2$, also denoted by $w$, on the hyperelliptic
surface $M$. This function defines a two-fold covering of the sphere 
branched over $2g+2$ points $Q_1,\ldots,Q_{2g+2}$, where $w(Q_i)=x_i$.
A basis for the space of holomorphic $1$-forms on $M$
is then given by
$$ \omega_k  \ := \ \frac{\o^{k-1}d\o }{\sqrt{P(\o)}} \ \ \ \ 
   \mbox{ for } \ \ \ \ k = 1, \dots, g \ \ . $$
It will be convenient to work with a canonical homology basis 
$\{\gamma_k,\;\tilde{\gamma}_k\}$ on $M$ chosen as in Figure 2.
We denote by $\Omega_{kl}$ the period of the $1$-form $\omega_l$
along the cycle $\gamma_k$, i.e.,
$$\Omega_{kl} := \oint_{\gamma_k}\omega_l
                   = \oint_{\gamma_k}\frac{w^{l-1}dw}{\sqrt{P(w)}}\ .$$
The basis of holomorphic $1$-forms, $\{\zeta_k\}$, dual to the canonical
homology basis $\{\gamma_k,\;\tilde{\gamma}_k\}$ is defined by the 
equation
$$\oint_{\gamma_k}\zeta_l=\delta_{kl}\ .$$
In particular, we have the relation
$$\omega_k=\sum_{l=1}^g\ \Omega_{lk}\;\zeta_l\ ,$$
and the matrix $\Omega$ is invertible. The {\sl period matrix} $\tau$ 
is given by
$$\tau_{kl} := \oint_{\tilde{\gamma}_k}\zeta_l\ ,$$
and it is known to be symmetric and to have positive definite imaginary part.
The surface $M$ has an anti-holomorphic involution induced by complex 
conjugation on the complex plane. In terms of the functions $w$ and
$\omega=\sqrt{P(w)}$, it can be written as
$$(w,\omega)\rightarrow(\bw,\bar{\omega})\ .$$
This involution will be used to extend the theory from the upper half-plane 
to the lower half-plane while taking care of the boundary conditions on 
the real axis. There is a second holomorphic involution that interchanges
the two sheets of $M$ and that can be written as
$$(w,\omega)\rightarrow(w,-\omega)\ .$$
This involution is used when passing from the sphere with cuts to its 
cover $M$. 
\mn\mn
\subsection{3.2\ \  The Ward identities} 
To begin with, we introduce the correlations that we plan to investigate 
below. Besides the primary bulk fields $\phi_\rg (z,\bar z) :=  
\exp(i \rg  X(z,\bar z))$,  they involve $2g+2$ boundary twist fields 
inserted at the points $x_i$, which induce changes between Dirichlet and 
Neumann boundary conditions. From our discussion in the previous section 
we know that there is an infinite number of boundary condition changing 
operators that we could insert. But it is sufficient to study the fields 
$\s(x)$ of conformal weight $h={1\over16}$, since all others can be 
obtained out of $\s(x)$  by OPE with chiral fields. Thus, our discussion 
deals with correlators of the form 
\be G(\bfz,\bfx) \  = \ 
   \langle \phi_1(z_1,\bar z_1) \dots \phi_n(z_n,\bar z_n) 
   \ \s(x_1) \dots \s(x_{2g+1}) \rangle\ \ 
\labela\corrG
\ee\mkgho\corrGho
where we use the notation $\phi_\nu = \phi_{\rg _\nu}$, and where the 
boundary field $\s(x_{2g+2}) = \s(\infty)$ is absorbed in the notation 
$\langle \dots \rangle = \langle \sigma | \dots |0\rangle$, with 
$|0\rangle$ denoting the vacuum state. We could insert further boundary 
fields $\Psi_\rg (z)$. Their interpretation depends on whether they are 
inserted in an interval with Dirichlet or Neumann boundary conditions. 
In the former case, they could induce jumps in the Dirichlet parameters 
$\der^k_0$ if they are associated with open strings stretching between
branes at different positions. A boundary operator $\Psi_\rg (z)$
inserted in one of the Neumann intervals, on the other hand, creates 
an open string which has both ends on the same brane (a Euclidean 1-brane, 
in our case) and moves with some definite momentum along its world-volume.
\mn
As far as Ward identities are concerned, correlation
functions of such boundary fields are actually more fundamental, 
since one may split the bulk fields $\phi_\rg (z,\bz)$ into a
product of $\Psi_\rg (z)$ and $\Psi_{-\rg }(\bz)$. For this reason, 
most of our investigations below involve the correlation functions    
\be F(\bfz,\bfx) \  = \ 
   \langle \Psi_1(z_1) \dots \Psi_n(z_n) 
   \ \s(x_1) \dots \s(x_{2g+1}) \rangle\ \ \ ,
\labela\corr
\ee\mkgvo\corrvo
{}from which the correlators $G(\bfz,\bfx)$ can be reconstructed. 
\bn
Our analysis will make essential use of the Mittag-%
Leffler theorem, and hence it is based on the study of singularities 
in correlation functions. The latter are encoded in the operator 
product expansions between chiral fields and the bulk and boundary 
fields appearing in \corrGho,\corrvo. For the Virasoro field $T$  
one has the standard expansions: 
\ba{ 
\cT(\o) \, \Psi_\rg (z)  & \sim \ \left[\, {h_\rg \over (\o-z)^2} +  
   {1\over \o-z}\, \partial_z\  \right] \ \Psi_\rg (z)\ \ , 
 \labelb\tpsi\cr\mkgho\tpsiho
\cT(\o)\  \sigma(x)  & \sim \ \left[ \,  {h_\sigma \over (\o-x)^2} 
 + {1\over \o-x }\, \partial_x \, \right] \ \sigma(x) \ \ .
\labelb\tsigma\cr\mkgvo\tsigmavo} 
\ea 
Here and in the following, the symbol $\sim$ means ``equal up to 
terms which are regular as $\o \rightarrow z$''. According to
the rule $\phi_\rg (z,\bz) \approx \Psi_\rg (z) \Psi_{-\rg }(\bz)$, 
the operator product expansions of $\cT$ with $\phi_\rg $ contain 
further terms in which $z$ is replaced by $\bar z$ (note that 
$h_\rg  = h_{-\rg }$). These formulas may be compared with 
equations \TP\ and \Tv\ in Section 2. For our correlation functions, 
eqs.\ \tpsiho,\tsigmavo\ imply
\ba{ & \hskip-4mm \langle \cT(\o)  \, \Psi_1(z_1) 
    \dots \Psi_n(z_n) 
   \ \s(x_1) \dots \s(x_{2g+1}) \rangle \ \bcr
      &  \hskip-2mm = \left[ \ \sum_{\nu = 1}^n \left(
      {h_\nu \over (\o-z_\nu)^2} + {1 \over \o - z_\nu} \, 
     {\partial \over \partial z_\nu} \right) + \sum_{i=1}^{2g+1} \left( 
     {h_{\sigma}  \over (\o-x_i)^2} + {1 \over \o - x_i} \, 
     {\partial \over \partial x_i} \right) \right]\ 
     F(\bfz,\bfx) \ . \labelb\Tward \cr}\ea
\mn
The situation is more subtle for the current $\cJ(\o)$, 
which we recall is only well-defined on a surface of genus
$g$. More precisely, $\cJ(\o) d\o$ is a meromorphic 1-form on 
a hyperelliptic surface. For the operator product expansion
of $\cJ(\o)$ with the field $\Psi_\rg (z)$, we shall use  
$$ 
\cJ(\o) \, \Psi_\rg (z) \, d\o \ \sim \  {\rg  \over \sqrt{P(\o)}}
{\sqrt{P(z)} \over \o - z} \ \Psi_\rg (z)\; d\o\ \ .  
\labela\Jopep$$\mkgho\Jopepho
Indeed, the right hand side has a first order pole at $\o = z$ 
with residue $\rg $ and is regular otherwise. Equation \Jopep\ 
generalizes  formula \JP\ in Section 2.2. To determine the 
operator product expansion between $\cJ$ and the twist field
$\s(x)$, we observe that the leading contribution  
$$ \cJ(w) \, \s(x) \ \sim\  (w-x)^{h_\tau- 1 -h_\s}\   
          \tau(x) + \dots
$$
must involve a field $\tau$ of conformal weight $h_\tau = 
h_\s + 1/2 + \Z$. Otherwise, the expansion would not be consistent 
with the periodicity properties of $\cJ$ close to the branch 
point at $w=x$. Among the boundary condition changing operators,
there is one field with conformal weight $h_\tau = \oh + {1\over16}$ 
which gives rise to the most singular contribution diverging 
with $(w-x)^{-1/2}$, cf.\ the spectrum \ZND\ computed above. 
\hbn
After multiplication of the previous equation with $dw$, 
we are supposed to study the right hand side in the 
local coordinate $ \xi = \sqrt{w-x}$. The outcome is rather 
simple: The form $(w-x)^{-1/2} dw = 2 d\xi$ on the right hand 
side is regular at $\xi =0$ so that we conclude      
\be \cJ(\o) \, \s(x) \, d\o \ \sim \ 0 \ \ , 
\labela\Jopes\ee\mkgvo\Jopesvo
i.e., the singular part of the operator product expansion between 
$\cJ(w)dw$ and the twist field $\s(x)$ vanishes. 
\bn
As we will see, the following important formula is a consequence of 
these operator product expansions: 
\ba{ & \hskip-2cm \langle \cJ(\o) \, \Psi_1(z_1) \dots \Psi_n(z_n) 
   \ \s(x_1) \dots \s(x_{2g+1}) \rangle \ \bcr
      &  = \ \left[\ 
   \sum_{\nu=1}^n \ {\rg _\nu \over \sqrt{P(\o)}} {\sqrt{P(z_\nu)}
   \over \o - z_\nu }\, + \sum_{k = 1}^{g} 
   {\alpha_k (\bfz,\bfx) \, \o^{k-1} \over \sqrt{P(\o)} } \ \right]\ 
   F(\bfz,\bfx)  
    \ \ , \labelb\Jinsert \cr}
\ea 
where 
$$
\alpha_k (\bfz,\bfx) 
  \ = \  \sum_{l=1}^g \Omega^{-1}_{kl}\, \left(\, i\, \Delta^k 
    \der_0 -  \sum_{\nu = 1}^n \rg _\nu B^k (z_\nu) \, \right) 
   \ \ ,
$$
and the parameters $\Delta^k \der_0 := \der_0^{k} - \der_0^{k+1}$ 
are obtained from the values of the bosonic field at the boundary. 
\sn
To derive the formula \Jinsert\ we exploit the fact that a  
meromorphic 1-form on a compact Riemann surface is determined by its 
principal part up to some holomorphic 1-from. 
Operator product expansions, on the other hand, contain all 
information about the principal part. Hence, from eqs.\ \Jopepho,\Jopesvo\  
we conclude that 
\ba{  & \hskip-1cm 
\langle \cJ(\o) \, \Psi_1(z_1) \dots \Psi_n(z_n) 
   \ \s(x_1) \dots \s(x_{2g+1}) \rangle \ \bcr 
& \hskip1cm = \ 
  \sum_{\nu=1}^n \ {\rg _\nu \over \sqrt{P(\o)}} {\sqrt{P(z_\nu)}
   \over \o - z_\nu } \, F(\bfz,\bfx)  \ + \ \sum_{k = 1}^{g} 
   {\beta_k (\bfz,\bfx) \, \o^{k-1} \over \sqrt{P(\o)} } \ . 
\labelb\Jins\cr}\ea
Note that the insertion points $z_\nu$ and $x_i$ parametrize 
a whole family of meromorphic 1-forms, and the coefficients
$\beta_k$ may depend on them. Actually, we can determine this 
dependence completely. To this end we integrate the above equation 
along a loop $\c_k$ which surrounds the interval $[x_{2k},
x_{2k+1}]$, as shown in Figure 2. On the right hand side of our 
equation, this integral may be expressed in terms of the matrix $\Omega$
and
$$ B^k (z) \ :=\  \oint_{\c_k} { 1\over \sqrt{P(w)}} 
   {\sqrt{P(z)} \over w - z } \ d\o \ \ . 
\eq\OmBdef$$
Note that the matrix elements $\Omega_{kl}$ and the functions 
$B^k(z)$ depend on the insertion points $x_i$. With these 
conventions we find 
$$ \oint_{\c_k} d\o\ ({\rm r.h.s.\ of\ \Jins}) \ = \ 
   \sum_{\nu =1}^n \, \rg _\nu \, B^k(z_\nu)\,  F(\bfz,\bfx) 
   + \sum_{l=1}^g \, \Omega_{kl} \, \beta_l \ \ . $$
Next, let us analyze the integral over the left hand side of  
equation \Jins. By a deformation, we can make the integration contour 
$\c_k$ symmetric under a reflection $\c \rightarrow \bar \c$ 
along the real line. Now, we may split $\c_k$ into two parts 
$\c^>_k, \c^<_k$ with the property $\Im \c^>_k \geq 0$ and 
$\Im \c^<_k \leq 0$, so that each piece lies entirely in one of 
the half-planes. With these conventions, our contour can be written 
as a composition $\c_k = \c^>_k \circ \c^<_k$ which obeys 
$\bar \c^<_k = - \c^>_k$. If we recall, in addition, that the field 
$\cJ$ coincides with $J$ on the upper and with $-\bJ$ on the lower 
half-plane we deduce 
$$ \oint_{\c_k} \cJ(w)\  d\o \ = \  \int_{\c^>_k} J(w) \, 
   d\o + \int_{\c^>_k} \bJ(\bar w)\, d\bar\o \ = \ 
    i\, \int_{\c_k^>} d X(w,\bar\o) \ = \ i\, (\der_0^{k}- 
    \der_0^{k+1}) 
  \ \ . $$
In the penultimate step, we have expressed the currents through 
the bosonic field $X$ by $J(w) = i\, \partial X(w,\bar w)$ and 
$\bJ(\bar w) = i\, \bar \partial X(w,\bar w)$. The contour 
integral over the differential $dX$ is finally determined by 
the values of $X$ at the boundary. For the integration 
over the left hand side of \Jins, this result implies that 
$$ \oint_{\c_k} d\o\ ({\rm l.h.s.\ of\ \Jins}) \ 
   = \ i\, (\der_0^{k}- \der_0^{k+1})\, 
   F(\bfz,\bfx) \ =: \ i\, \Delta^k \der_0\,  F(\bfz,\bfx) \ \ . $$
Putting all this together, we arrive at the following formula
for the function $ \beta_k(\bfz,\bfx)$: 
$$ \beta_k(\bfz,\bfx) \ = \ \sum_{l=1}^g \, 
   \Omega^{-1}_{kl} \left( i\, \Delta^l \der_0 - \sum_{\nu=1}^n
   \rg _\nu B^l (z_\nu) \right) \ F(\bfz,\bfx) \ =: \ 
   \alpha_k(\bfz,\bfx) \ F(\bfz,\bfx)  \ \ . $$
The functions $\alpha_k$ introduced here depend on the insertion 
points $z_\nu, x_i$, the charges $\rg _\nu$ and the values $\der^k_0$ 
of the bosonic field at the boundary. Additional information, 
e.g., on the unknown function $F(\bfz,\bfx)$, is not needed. 
This concludes our derivation of eq.\ \Jinsert. 
\bn\bn\bn
\vbox{
\hbox{\hskip-1.1cm\epsfbox{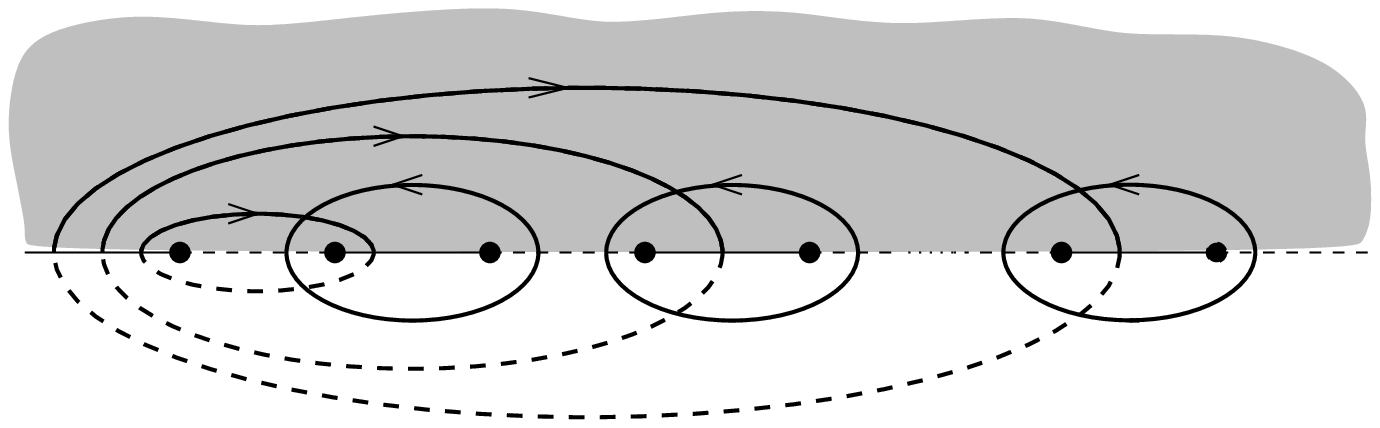}}
\sn
\vskip-4.2cm\hbox{\hskip8cm$\tilde\gamma_g$}
\vskip.15cm\hbox{\hskip5.9cm$\tilde\gamma_2$}
\vskip.15cm\hbox{\hskip2.5cm$\tilde\gamma_1$}
\vskip-.3cm\hbox{
  \hskip3.55cm\hskip.75cm$\gamma_1$
  \hskip1.95cm\hskip.75cm$\gamma_2$\hskip2.85cm\hskip.8cm$\gamma_g$}
\vskip.25cm\hbox{\hskip1.95cm$x_1$\hskip.8cm$x_2$\hskip1.25cm$x_3$
\hskip1.3cm$x_4$\hskip1cm$x_5$\hskip2.3cm$x_{2g}$\hskip.65cm$x_{2g+1}$}
\vskip1.9cm 
\mn
\narrower
{\bf Figure 2:} The curves $\gamma_k$ run counterclockwise around the 
Neumann cuts on the real line. The curves $\widetilde{\gamma}_k$ run 
clockwise and close on the second sheet through Neumann intervals.
}
\bn\mn
\subsection{3.3\ \ The twisted Knizhnik-Zamolodchikov equations}  
We now start to exploit formula \Jinsert\ together with the Sugawara 
construction of the energy-momentum tensor $\cT$ to compute
the effect of inserting the field $\cT$ into our correlation 
functions. If we include one extra field $\Psi_\rg (u)$ into eq.\ 
\Jinsert, differentiate with respect to $u$ and set $\rg  = 0$, 
we obtain as a consequence of $\cJ(u) = \frac{1}{\rg } \partial_u 
\psi_\rg(u)  |_{\rg =0}\,$ 
\ba{ & \hskip-4mm \langle \cJ(\o) \, \cJ(\u) \, \Psi_1(z_1) 
    \dots \Psi_n(z_n) 
   \ \s(x_1) \dots \s(x_{2g+1}) \rangle \ \bcr
      &  = \ \left[\ 
   \sum_{\nu,\mu =1}^n \ {\rg _\nu \rg _\mu \over \sqrt{P(\u)} \sqrt{P(\o)}} 
    {\sqrt{P(z_\nu)} \over \o - z_\nu }{\sqrt{P(z_\mu)} \over \u - z_\mu }
    + \sum_{k = 1}^g \sum_{\nu=1}^n   
   {\alpha_k (\bfz,\bfx) \, \o^{k-1} \over \sqrt{P(\u)} \sqrt{P(\o)} } 
   {\rg _\nu \sqrt{P(z_\nu)} \over \o - z_\nu} \right. \bcr
   & \hskip7mm + \left. \sum_{k,l=1}^g {\alpha_k\alpha_l \o^{k-1} \u^{l-1} 
    \over \sqrt{P(\o)} \sqrt{P(\u)}} + 
    {1 \over \sqrt{P(\o)}} \, \frac{d}{d\u}\, \left( 
    {\sqrt{P(\u)} \over \o - \u } - \Omega^{-1}_{kl} 
     B^l(\u) \o^{k-1}\right)\, \right] \ F(\bfz,\bfx) \ \ . \cr} \ea    
At this stage we can subtract the term $1/(\o-\u)^2$, multiply by 
a factor $1/2$ and perform the limit $\u \rightarrow \o$. A short and
elementary computation gives   
\ba{ & \hskip-4mm \langle \cT(\o)  \, \Psi_1(z_1) 
    \dots \Psi_n(z_n) 
   \ \s(x_1) \dots \s(x_{2g+1}) \rangle \ \bcr
      &  = \ \left[\  \frac12
   \sum_{\nu,\mu =1}^n \ {\rg _\nu \rg _\mu \over P(\o) } 
    {\sqrt{P(z_\nu)} \over \o - z_\nu }{\sqrt{P(z_\mu)} \over \o - z_\mu }
    +  \sum_{k = 1}^g \sum_{\nu = 1}^{n}  
   {\alpha_k (\bfz,\bfx) \, \o^{k-1} \over P(\o) } 
   {\rg _\nu \sqrt{P(z_\nu)} \over \o - z_\nu} \right. \bcr
   & \hskip5mm + \left. \frac12 \sum_{k,l=1}^g {\alpha_k\alpha_l \o^{k+l-2} 
    \over P(\o)} - 
    \, {\sqrt{P(\o)}'' \over 4 \sqrt{P(\o)}} 
     -  \frac12 \sum_{k,l=1}^g { \Omega^{-1}_{kl} 
     B^l(\o)' \o^{k-1} \over \sqrt{P(\o)}} \, \right] 
   \ F(\bfz,\bfx) \ \ . \cr} \ea
The same correlator has been computed in eq.\ \Tward\ directly with 
the help of operator product expansions between the Virasoro field 
$\cT$ and $\Psi_\rg $, $\sigma$ from \tpsiho,\tsigmavo. Comparison 
of the residues at $\o = z_\nu$ in the two different expressions gives 
the $z$-components of the Knizhnik-Zamolodchikov equations:  
$$ \partial_{z_\nu} F(\bfz,\bfx) \ = \ \left[\, 
   \sum_{\mu \neq \nu}^n \ {\sqrt{P(z_\mu)} \over \sqrt{P(z_\nu)}}  
    {\rg _\nu \rg _\mu  \over z_\nu - z_\mu }  - 
    \sum_i \, \frac{h_\nu}{z_\nu - x_i} + \sum_{k = 1}^{g}  
   {\rg _\nu \alpha_k \, z_\nu^{k-1} \over \sqrt{P(z_\nu)}}\, 
   \right]\ F(\bfz,\bfx)  \ . $$
If we restrict to the case of two twist field insertions, i.e.\ $g=0$, 
the last term vanishes. Putting $x_1=0$ and hence $P(\o) = \o$, 
formula \simpleKZeq\ is recovered. 
\mn
Computation of the residues at $\o=x_i$ yields a formula for the 
derivative of $F$ with respect to the position $x_i$ of twist fields. 
Using that
$$
\Res_{x_i}  \left( \sqrt{P(\o)}'' \over 4 \sqrt{P(\o)}\right)
   \ = \ \frac{1}{8} \ \sum_{j \neq i} \ \frac{1}{x_i - x_j}
\phantom{xxx}$$
and 
$$   \Res_{x_i} \left( \sum_{k,l=1}^g { \Omega^{-1}_{kl} 
     B^l(\o)' \o^{k-1} \over \sqrt{P(\o)}} \, \right)
     \ = \ \sum_{k,l=1}^g \Omega_{kl}^{-1} \int_{\gamma_l} 
    \frac12 \frac{x_i^{k-1}}{\sqrt{P(\xi)}} \frac{1}{\xi-x_i}\ d\xi
$$
we obtain  
\ba{ \partial_{x_i} F(\vec z,\vec x) & \ = \ \left[\ 
         \frac{1}{2\prod_{j\neq i}
           (x_i -x_j)} \left(\;\sum_{\nu =1}^{n}  
       \frac{\rg_\nu \sqrt{P(z_\nu)}}{x_i-z_\nu}   
        +  \sum_{k=1}^g \alpha_k(\vec z,\vec x)  
   \ x_i^{k-1} \right)^2 \right. \cr & \hskip5mm \left. 
      - \frac{1}{8}\ 
       \sum_{j \neq i} \frac{1}{x_i-x_j}  
   - \frac{1}{4}\ \sum_{k,l=1}^g \, \Omega^{-1}_{kl} \oint_{\gamma_l}
   {\frac{x_i^{k-1}}{\sqrt{P(\xi)}(\xi-x_i)}}^{\phantom{\frac12}}
   d\xi\ \right]\  F(\vec z,\vec x)
    \ \ . \labelb\xdeq \cr} \ea
To conclude this section, we come back to the original correlators
$G(\vec z,\vec x)$ of bulk-fields $\phi_\rg(z,\bar z)$ and boundary
twist fields $\sigma(x_i)$. The differential equations they obey
will be formulated with the help of the following functions:         
\ba{ \omega_0(z,w) \ & = \ \frac{1}{\sqrt{P(z)}} \left[ \frac{\sqrt{P(w)}}
 {z-w} - \frac{\sqrt{P(\bw)}}{z-\bw} \right] - 
 \sum_{k,l=1}^{g} \frac{z^{k-1}}{\sqrt{P(z)}}\; \Omega^{-1}_{kl} 
  \left( B^l(w) - B^l(\bw)\right)\ , \bcr 
 \lambda_0(z) \ & = \ - \oh\;\sum_{i=1}^{2g+1} \frac{1}{z-x_i} 
 - \frac{1}{\sqrt{P(z)}} \frac{\sqrt{P(\bz)}}{z-\bz} - 
 \sum_{k,l=1}^g \frac{z^{k-1}}{\sqrt{P(z)}}\; \Omega^{-1}_{kl} 
 \left(B^l(z)- B^l(\bz)\right)\ , \bcr 
 \sigma_\Delta(z) \ & = \ \sum_{k,l=1}^g \frac{z^{k-1}}{\sqrt{P(z)}}\; 
  \Omega^{-1}_{kl}\, \Delta^l \der_0 \ \ . \cr} \ea 
Taking into account the equation $\phi_\rg(z,\bar z) =
\Psi_\rg(z) \Psi_{-\rg} (\bz)$, we conclude that the correlation
functions $G(\vec z,\vec x)$ must satisfy the following set of
first order linear differential equations:
$$ \partial_{\xi}  G(\vec z,\vec x) \ = \
     A_{\xi} \ G(\vec z,\vec x)\ \ \ \mbox{ for\  all } \ \ \
     \xi = z_\nu,\bz_\nu,x_i 
\labela\diffeq $$
with the connection matrices $A_{z_\nu}$, $A_{\bz_\nu}$ and 
$A_{x_i}$ being defined by 
\ba{ 
A_{z_\nu} \ & = \ \sum_{\mu \neq \nu} \rg_\nu \rg_\mu 
  \omega_0(z_\nu, z_\mu) + \rg_\nu^2\lambda_0(z_\nu) + i g_\nu 
  \sigma_\Delta(z_\nu)\ \ ,  \labelb\Acon\mkgho\Aconho
\cr 
A_{\bz_\nu} \ & = \ \sum_{\mu \neq \nu} \rg_\nu \rg_\mu 
  \overline{\omega_0(z_\nu, z_\mu)} + \rg_\nu^2\overline{\lambda_0(z_\nu)}
 + i g_\nu \overline{\sigma_\Delta(z_\nu)} \ \ ,
  \labelb\bAcon\mkgvo\bAconvo\mkgvho\bAconvho 
\cr     
A_{x_i} \ & = \ \frac12 \lim_{x \rightarrow x_i} (x-x_i) \left(
   \sum_{\nu = 1}^n \rg_\nu \omega_0(x,z_\nu) +
   i \sigma_\Delta (x) \right)^2 \bcr & \hskip2cm
   - \frac{\partial_i H_i(x_i)}{8 H_i(x_i)} 
   - \frac{1}{4}\  \sum_{k,l=1}^g \,\Omega^{-1}_{kl} \oint_{\gamma_l}
   {\frac{x_i^{k-1}}{\sqrt{P(\xi)}(\xi-x_i)}}^{\phantom{\frac12}}
   d\xi \ \ .\labelb\Axcon\mkgvo\Axconvo
\cr}\ea
We have introduced the function $H_i(x_i) := \prod_{j\neq i}
(x_i - x_j)$. Note that the term in brackets has a simple
pole at $x= x_i$ which we cancel by the extra factor $x-x_i$ before
performing the limit. Equations \diffeq\ through \Axcon\
constitute the main result of this section. 
    
\bn
\section{4.\ \ Construction of correlation functions}%
\def\bv{{\bar v}} 
It remains to reconstruct the correlation functions $G(\vec z,
\vec x)$ from the system of linear first order differential equations
that we obtained in the previous section. Integration of the
equations is, in principle, straightforward, but it leaves one constant
factor undetermined. The latter is found explicitly in terms of the
boundary conditions. Moreover, we shall manage to express the correlators
$G(\vec z,\vec x)$ in terms of rather elementary building blocks.  
\bn

\subsection{4.1\ \   Integration of the {\fat z}-connection}
To begin with, we simplify our problem by fixing the insertion 
points of the boundary twist fields $\sigma(x_i)$  and
considering only the dependence of $G_{\vec x}(\vec z) =
G(\vec z,\vec x)$ on the positions $(z_\nu,\bz_\nu)$ of bulk
fields. This means that we have to integrate the $z$-connection
$A_{z_\nu} dz_\nu + A_{\bz_\nu} d\bz_\nu$ defined in eqs.\ 
\Aconho,\bAconvo\ . The result will be written with the help of two 
functions $G_0(z,w)$ and $S_0(w)$ which are given by 
\ba{  & G_0(z,w)  \ := \ 2\ \Re \int_x^z \ \omega_0(\xi,w) \, d\xi\ \ ,
     \labelb\Gfun\mkgho\Gfunho \cr
     S_0(z) & \ := \ \lim_{v \rightarrow x}\left[^{\phantom{\frac12}} 
     2 \ \Re \int_v^z \ \lambda_0(\xi) \, d\xi - \log(v-\bv) - 
     \Re \log P(v) \ \right]\ \ .  \labelb\Sfun\mkgvo\Sfunvo \cr} 
\ea 
Here, the point $x$ is chosen to lie in the Dirichlet-interval $]x_1,x_2[$.
The integrand in eq.\ \Gfun\ is regular on the real axis and hence the
integral is well defined. One can easily see that its value neither 
depends on the starting point $x \in \; ]x_1,x_2[$ nor on the choice of the 
curve $\gamma$ from $x$ to $z$. 
In contrast, the integrand $\lambda_0(\xi)$
in our definition of $S_0(z)$ has a simple pole on the real axis. This
is the reason why we subtract the divergent term $\log(v-\bv)$ 
before taking the limit $v \rightarrow x$. The contribution $-\Re \log P(v)$
is added to render the whole function independent of the integration
contour and, in particular, of the starting point $x$. More details
on the definition of $S_0(z)$ and a discussion of its properties can 
be found in Appendix B. 

\mn
The two functions we have just defined possess a number of abstract 
properties that characterize them uniquely. First of all, it is not 
difficult to see that $G_0(z,w)$ is simply a Green's function on 
the upper half-plane, i.e., it obeys   
$$
\Delta_z G_0 (z,w)  \ = \ 4 \pi \delta(z-w)  \quad \ \
     \mbox{ for } \ \ \ \Im z > 0 
$$
and the boundary conditions 
$$     G_0(z,w)  \ = \ 0 \ \ \ \mbox{ for }
      \ \ \ z \  \in\ D_i \ \ \  , \quad \quad \  
     \frac {\partial} { \partial \Im z}\  G_0(z,w)  \ = \  
     0 \ \ \ \mbox{ for } \ \ \ z \ \in \ N_i \ \ .
$$
A standard computation shows that $G_0(z,w)$ is symmetric in its 
arguments, $G_0(z,w) = G_0(w,z)$.
\sn
The function $S_0(z)$, on the other hand, is harmonic throughout
the whole upper half-plane, i.e.\ $\Delta_z S_0(z) = 0$. 
It diverges at the boundary with a leading singularity of the form
$$
     S_0(z)  \ \sim \ \mp \, \log |z-\bz| \ + \ \dots \ \ \ \mbox{ for } 
      \ \ \ \Re\ z \ \in\ \ \ \biggl\{ {D_i \atop N_i}  \ \ .
\eq\singS$$ 
We are now in a position to integrate the differential equations
\Aconho,\bAconvo\  for the correlators of bulk fields $\phi_\rg(z,\bar z)$.
The result is given by 
\ba{ & \log  G_{\vec x} (\vec z)  \ = \ \int_{\vec w}^{\vec z}  (A_{z_\nu}
    d\xi_\nu 
  + A_{\bz_\nu} d\bar \xi_\nu) \quad +\ \; \Lambda(\vec w) \bcr 
 & \ = \ \frac{1}{2}\sum_{{\nu,\mu=1\atop \nu\neq\mu}}^n \rg_\nu \rg_\mu \, 
 G_0(z_\nu,z_\mu) 
  + \sum_{\nu=1}^n \rg_\nu^2 \, S_0(z_\nu) +  
 \sum_{\nu = 1}^n \,\frac{i \rg_\nu}{4 \pi} \;\sum_{i=1}^{g+1} \int_{D_i}
  \der_0^i \,\frac{\partial}{\partial 
  \Im \xi}\  G_0(\xi,z)\ d\xi \  .\phantom{xxx} \labelb\lnG 
   \cr } 
\ea     
In the first line we have chosen some arbitrary curve in the
configuration space of $n$ particles in the upper half-plane
starting at points $w_\nu$ with $\Im w_\nu > 0$. The
integration ``constant'' $\Lambda(\vec w)$ depends on the
choice of the starting point and has to be fixed such that
the resulting function $\log G_{\vec x}(\vec z)$ satisfies the
desired boundary conditions. In passing to the second line, we
have extended the integration to $w_\nu = x$ on the boundary and
inserted the definitions \Gfunho,\Sfunvo\ of the functions
$G_0$ and $S_0$. Then we use the auxiliary formula
$$ \sum_{i=1}^{g+1}\, \frac{1}{4\pi} \;\int_{D_i} \der_0^i
  \frac{\p}{\p \Im \xi} \ G_0(z,\xi) \, d\xi \ = \
  2\Re \sum_{k=1}^g \int_x^z \frac{\xi^{k-1}}{\sqrt{P(\xi)}}
  \ \Omega_{kl}^{-1} \Delta^l \der_0 \;d\xi \ + \ \der_0^1\ \ .  $$
To prove this formula one should notice that the function on the l.h.s.\  
of the equation is harmonic in the upper half-plane, that it satisfies 
Neumann boundary conditions along the intervals $N_k$ and that it approaches 
the constant values $\der_0^k$ for $z \in D_k$. By explicit computation,  
one can establish the same behaviour for the function on the r.h.s. 
Since these properties are sufficient to determine the functions 
uniquely, the desired equation follows. We employ it to bring the third 
term of eq.\ \lnG\ into a form that allows the most explicit control of 
the boundary behaviour and hence is quite appropriate for fixing the 
remaining $\Lambda(x)$.
\bn 
        
\subsection{4.2\ \  Integration of the {\fat x}-connection} 
Before we address the integration of the full connection
\Aconho,\bAconvho,\Axconvo\ below, we investigate another
simplified situation in which there are no bulk fields
present. Consequently, the charges $g_\nu$ in the equation
\xdeq\ can be set to zero, and we are confronted with the problem
of solving the following equation for $Z_{\Delta}(\vec x)
:= G(\vec{x})$:  
$$ 2 H_i \,\partial_i \log Z_{\Delta}(\vec x) \ = \ 
   \left( \sum_{k=1}^g \Omega^{-1}_{kl} i \Delta^l \der_0 
   \ x_i^{k-1} \right)^2 - \frac{1}{4}\ \partial_i H_i 
   - \frac12\ \sum_{k,l=1}^g \,\Omega^{-1}_{kl} \oint_{\gamma_l}
   \frac{x_i^{k-1} H_i}{\sqrt{P(\xi)}(\xi-x_i)} \ d\xi 
    \ , $$ 
where $H_i$ denotes the function $H_i = \prod_{j \neq i} 
(x_i - x_j)$ as before, and $\Delta=\{\Delta^l \der_0\}$.
These differential equations were solved by Zamolodchikov in \q{\Zamo}. 
Here we simply quote the final result: 
$$ 
Z_{\Delta}(\vec x) \ = \prod_{i>j}^{2g+1} (x_i - x_j)^{-{1\over8}} 
   \; {\det}^{-\oh}( \Omega) \; e^{\frac{i}{8 \pi} \sum_{k,l=1}^g 
   \Delta^k \der_0 \Delta^l \der_0 \, \tau_{kl}} \ .
$$ 
\mn

\subsection{4.3\ \  The correlation 
functions \hbox{\fat G}($\vec{\hbox{\fat z}},\vec{\hbox{\fat x}}$)}
The results of the previous two subsections can be combined 
into explicit expressions for the correlators $G(\vec z,\vec x)$ 
of bulk fields $\phi_\rg(z,\bar z)$ and boundary twist fields 
$\sigma(x)$. To see this we note that solutions of the 
differential equations \diffeq\  can be found by integrating
the connection 1-form $A_{w_\nu} dw_\nu + A_{\bar w_\nu} d\bw_\nu +
A_{\xi_i} d\xi_i$ along an arbitrary curve $\gamma(t) = (\gamma_{\vec z}(t),
\gamma_{\vec x}(t)),\ t \in [0,1],$  that ends at the point $(\vec z,
\vec x)$ in the $(2n+2g+1)$-dimensional real configuration space. 
\hbn
With some care (see the first subsection) we can 
start the integration with all insertion points being on the real axis.
Furthermore, we may choose $\gamma$ such that all twist-fields
are moved to their final position at $\vec x$ {\sl before} we
begin moving the bulk fields into their desired locations. 
\hbn
In more mathematical terms this means that $\gamma$ consists
of two parts $\gamma = \gamma^{(2)} \circ \gamma^{(1)}$ with
$\partial_t \gamma^{(1)}_{\vec z}(t\in [0,\frac12]) = 0$ and
$\partial_t \gamma^{(2)}_{\vec x}(t\in [\frac12,1]) = 0$. As
long as $t \leq \frac12$, the bulk fields are located at points
$w_\nu = \gamma_\nu(t)$ belonging to the first Dirichlet interval 
$D_1 = ]x_1,x_2[$. This implies $\omega_0(x,\gamma_\nu(t)) = 0$ for 
$t \in [0,\frac12]$, so that one term in our expression \Axcon\ for 
$A_{x_i}$ drops out. Hence, we are precisely in the situation considered
in the previous subsection, and the integration of our connection
1-form over $\gamma$ in the interval $t \in [0, \frac12]$ gives
$Z_{\Delta}(\vec x)$. When we continue the integration to $t=1$, 
we add the expression for $\log G_{\vec x} (\vec z)$ computed in 
eq.\ \lnG. After some rewriting, the following result for the 
correlations function $G(\vec z,\vec x)$ is obtained:  
\be G(\vec z,\vec x) \ = \ 
   Z_{\Delta}(\vec x) \ \exp \left(\sum_{\nu=1}^n \rg_\nu^2 S_0(z_\nu)
     \right) 
    \ \exp \left(\sum_{\nu =1}^n \, i \rg_\nu \Phi^{(n-1)}_{\der_0}
   (z_\nu) \right) \ \ .\labela\corr \ee 
Here, $\Phi^{(n-1)}_{\der_0} (z_\nu)$ denotes the potential that 
is created by $n-1$ charges $\rg_\mu$ at points $z_\mu \neq z_\nu$ 
in the presence of the boundary with mixed boundary conditions,  
$$ \Phi^{(n-1)}_{\der_0} (z_\nu) \ = \ \frac{1}{2}
   \int_{\Im w > 0} \hskip-3mm
   d^2w\ G_0(z_\nu,w) \sum_{\mu\neq \nu} \rg_\mu\, \delta(w-z_\mu)
     + \frac{1}{4\pi}\, \sum_{i=1}^{g+1} \; \int_{x_{2i-1}}^{x_{2i}}
     \hskip-2mm  \der_0^i \, \frac{\partial}{\partial \Im \xi}\
     G_0 (\xi,z_\nu) \ d\xi \ \ . 
$$        
It is quite instructive to interpret each of the three factors in our 
final expression for the correlation function $G(\vec z,\vec x)$
directly within conformal field theory. For the moment, let us specify 
the number of bulk fields in $G$ by some extra superscript, i.e.\ we 
shall write $ G(\vec z,\vec x) = G^{(n)} (\vec z,\vec x)$.
Now consider the object
$$ \Phi(z_\nu)\ := \ \frac{1}{i \rg_\nu} \log \left(\
        \frac{G^{(n)} (\vec z,\vec x)}{\exp(\rg^2 S_0(z_\nu))
       \,  G^{(n-1)}(\vec z',\vec x)} \ \right)    
\ \ \  
$$
where $\vec z'$ denotes the set of $n-1$ bulk coordinates 
$z_\mu,\ \mu \neq \nu$. It is easy to determine the behaviour of 
$\Phi(z_\nu)$ as a function of the bulk coordinate $z_\nu$ from the 
bulk and bulk-boundary operator product expansions of the fields 
$\phi_\rg(z_\nu,\bz_\nu)$, cf.\ Subsect.\ 2.3:  
\ba{  & \Delta_{z_\nu}\, \Phi(z_\nu)  \ = \ 2 \pi \
     \sum_{\mu \neq \nu} \ \rg_\mu 
                         \ \delta(z_\nu-z_\mu) \bcr
    \Phi(z_\nu)  \  = \ \der_0^i \ \ \ & \mbox{ for }
      \ \ \   z_\nu \ \in\ D_i \ \ , \ \quad  \  
     \frac {\partial} { \partial \Im z_\nu}\  \Phi(z_\nu)  \ = \  
     0 \ \ \ \mbox{ for } \ \ \ z \ \in \ N_i 
\ \ . \ \  \cr}
\ea     
These properties characterize the function $\Phi(z_\nu)$ uniquely, and 
by standard formulas from electrostatics we obtain that $\Phi(z_\nu) =
\Phi^{(n-1)}_{\der_0} (z_\nu)$. An iteration of this construction
along with $G^{(0)}(\vec x) = Z_{\Delta}(\vec  x)$ leads to our product 
formula \corr\ for the correlation function $G(\vec z,\vec x)$.  
\hbn
The three factors can be interpreted as follows: $Z_{\Delta}(\vec  x)$ 
is a ``partition function'' corresponding to some line charge distribution 
provided by the twist fields alone; the term
$$\sum_{\nu=1}^n\  g_{\nu}\; \Phi^{(n-1)}_{\der_0}(z_\nu)$$
is the electrostatic potential corresponding to a configuration
of $n$ point particles with charges $g_1,\ldots,g_n$
located at the points $z_1,\ldots,z_n$ and line charges distributions along
the Dirichlet intervals. The term
$$\sum_{\nu=1}^n\ g_{\nu}^2\; S_0(z_\nu)$$
can be interpreted as a renormalized electrostatic self-energy of the 
point particles located at $z_1,\ldots,z_n$. 
\bn
\subsection{4.4\ \ Path integral approach and extensions} 
Before we conclude, let us briefly sketch how the theory can be 
formulated in the path integral approach. This is important for the 
following two reasons: First, as long as one is only interested in 
free bosons, the path integral approach is a powerful alternative to 
our analysis above involving Knizhnik-Zamolodchikov connections. The 
path integral formulation allows for a more direct computation of 
the correlation functions (but it does not easily extend to non-%
abelian group targets.) Secondly, using path integrals we will be 
able to describe rather easily possible extensions of our analysis 
to compact targets and to D-branes with B-fields. 
\hbn 
To begin with, we consider once more the familiar situation of a 
non-compact 1-dimensional target and Dirichlet parameters $\der_0^i$. 
We denote by $G$ the Green's function of the Laplacian on the upper 
half-plane,
$$\Delta_z G(z,w)\  = \ \delta(z-w)\ ,$$
subject to the boundary conditions
\ba{ G(x,w) &=\ 0\ \quad {\rm for}\ x\in D\ , \cr
     \partial_y \, G(x,w)&=\ 0\ \quad{\rm for}\ x\in N\ . 
\cr}\ea
The Gaussian measure with covariance $G$ and mean $0$ is denoted by $\mu_G$,
and we use $\chi$ for the corresponding random variable. With the help of 
the Dirichlet parameters $\der_0^i$, we define a real-valued function 
$\xi$ on the upper half-plane by 
\ba{\Delta\xi\  & =\ 0\ ,\cr
    \xi|_{D_i}\  =\ \der_0^i\ & ,\quad \partial_y
   \xi|_{N_i}\ =\ 0 \ ,\quad i\, =\, 1,\ldots,g+1\ .\cr
}\ea            
The effect of the Dirichlet parameters is incorporated through a 
shift of the random variable $\chi$ by $\xi$ which gives us the 
bosonic field $X= \chi + \xi$. It appears when we construct the 
basic fields of the theory, namely the vertex operators
$$
\varphi_g(z,\bz)\ =\ {\bf :}\,e^{igX(z,\bz)}{\bf :}\;\ .
$$
In this framework we could recover the correlation functions above by 
integrating products of fields $\varphi_g(z,\bz)$ using the Gaussian 
measure. 
\bn
When the free boson $X$ takes values in a circle $S^1$, the boundary 
conditions depend both on Dirichlet parameters $\der_0^i$ and on Neumann 
parameters denoted by $\ter_0^i$. The Neumann parameters determine the 
Dirichlet parameters of the T-dual theory and can be thought of as the 
strength of constant Wilson lines turned on along a Neumann direction. 
In Sect.\ 3, we have worked in a local chart with the Neumann 
intervals cut out. Information on Neumann parameters is, therefore, 
lost unless we take a second chart into account which has cuts 
along the Dirichlet intervals. As above, we can derive 
Knizhnik-Zamolodchikov equations for each chart; the full correlation 
functions of the compactified theory are to be built up from the 
respective solutions in such a way that the boundary conditions are met.
\sn
The path integral computation of correlators is rather easy to 
adjust to the compactified situation: if we set Neumann parameters 
to zero and restrict attention to the fields $\varphi_g(z,\bz)$, 
we can use precisely the same formulas as above. 
\hbn
In order to incorporate non-vanishing Neumann parameters $\ter_0^i$
and to compute more general correlators for fields $\varphi^{(c)}_{g,
\bar{g}} (z,\bz)$ with $(g,\bar g)$ taken from an even, self-dual 
Lorentzian lattice, we introduce a real-valued function $\eta$ on 
the upper half-plane defined by
\ba{ \Delta\eta\ & =\ 0\ ,\cr     \eta|_{N_i}\ =\ \ter_0^i\ & ,
     \quad \partial_y\, \eta|_{D_i}\ =\ 0\ ,
        \quad i\,=\, 1,\ldots,g+1\ . \cr
}\ea
Then, we set
$$
\varpi_{g,z}(w)\ := \ g G'(w,z)\, +\, \eta(w)\ ,
$$
where $G'$ denotes the Green's function of the Laplacian on the upper 
half-plane with interchanged Dirichlet and Neumann boundary conditions, 
i.e.,
\ba{\Delta G'(w,z)&=\ \delta(w-z)\ ,\cr
    \partial_y\, G'(x,z)&=\ 0\ \quad{\rm for}\ x\in D\ ,\cr
    G'(x,z)&=\ 0\ \quad{\rm for}\ x\in N\ .\cr
}\ea
We now introduce the {\sl disorder operators} $D_g(z,\bz)$ satisfying
$$
D_g(z,\bz)F[dX]\ =\ F[dX+*d\varpi_{g,z}]\ ,
$$
for any functional $F$. The left- resp.\ right-moving chiral vertex 
operators can then be written as 
$$
\psi_g(z)\ =\ \varphi_{\frac{g}{2}}(z,\bz) D_{\frac{g}{2}}(z,\bz)\ ,\quad\quad
\bar{\psi}_{\bar{g}}(\bz)=\varphi_{\frac{\bar{g}}{2}}(z,\bz)
        D_{-\frac{\bar{g}}{2}}(z,\bz)\ ;
$$
see also \q{\FM} for more details and for an  application to 
soliton quantization in 2-dimensional theories. The basic fields of the 
compactified theory are products
$$
\varphi^{(c)}_{g,\bar{g}}(z,\bz)\ =\ \psi_g(z)\bar{\psi}_{\bar{g}}(\bz)
$$
where $(g,\bar{g})$ lies in some even, self-dual Lorentzian lattice.
For another approach to the rational compactified boson, the reader is 
referred to \q{\FSms}.
\bn\mn
Another extension would involve the appearance of $B$-fields
on our D-branes. This has attracted some interest recently, 
because of its relation with non-commutative geometry, see e.g.\ 
\q{\DH,\Vol,\SW, \ARStwo,\FFFSone} and references therein. 
Non-trivial B-fields can only exist if one of our branes is 
at least 2-dimensional. For simplicity, we shall focus on a 
pair of a Dp-and a D0-brane. The field strength on the Dp-brane 
will be denoted by $B$. In terms of boundary conditions for a 
multi-component free bosonic field, the situation is described 
as follows
$$ \partial_t X^a(t,0) \ = \ 0 \ \ \ \ \mbox{ and } \ \ \ \ 
 \partial_\sigma X^a(t,\pi) \ = \ B^a_{\, b}\ \partial_t X^b(t,\pi)
\ \ \ \  \mbox{ for } \ \ \ \ a,b = 1,\dots, p\ \ . $$
The spectrum of the associated boundary condition changing operators 
and the Green's functions in the presence of two twist fields have 
been discussed at various places (see e.g.\ \q{\SW,\CIMM}). Our techniques 
from Sections 3 and 4 allow to extend such investigations to the case of 
multiple twist insertions. Instead of giving the details here, we simply 
state how
one has to adjust the path integral computation to the new scenario. 
This is rather easy: All it requires is to replace the function $G$ 
above by some matrix valued Green's function $G_B = (G^{ab}_B)$.
The latter is a Green's function for the Laplacian ${\bf 1}_p\, \Delta$ 
on the upper half-plane (${\bf 1}_p$ denotes the p-dimensional 
identity matrix), subject to the boundary conditions
\ba{ G^{ab}_B(x,w) &=\ 0\ \ \ \ \ \ \ \quad {\rm for}\ x\in D\ , \bcr
   \partial_y \, G^{ab}_B (x,w)&=\ i B^{a}_{\, c}\  \partial_x
    G^{cb}_B(x,w) \ \quad{\rm for}\ x\in N\ . 
\cr}\ea  
With the help of this function, the calculation of correlators
proceeds as before.

\section{5.\ \  Outlook}%
We have succeeded in decomposing the complete bulk and boundary 
correlators in the presence of DN-transitions into functions 
with rather natural interpretations -- both from the point of view of 
electrostatics and from the CFT perspective. This is
useful for carrying out the remaining step in the computation 
of string amplitudes, namely the integration over insertion 
points of fields on the world-sheet. The calculation of such
string amplitudes gives effective actions involving a hyper-%
multiplet $\chi$ which comes with the twist fields. To leading 
order, the bosonic part of these actions can be found in \q{\Dou,
\Mal,\Hash}. Multiple twist insertions allow to compute higher 
order corrections. 
\sn
When we turn on a $B$-field, the string amplitudes may be 
described through field theories on some non-commutative space. 
It was suggested in \q{\SW} that these theories are related 
to some model on an ordinary commutative space through a 
complicated non-linear transformation. This statement can be 
checked order by order in the effective description. After the 
appropriate (but straightforward) extension to non-vanishing 
$B$-fields, the considerations presented above may be used to 
perform a similar analysis for theories which contain a  
hyper-multiplet $\chi$.   
\mn
Keeping the bulk insertions fixed, the sequence of correlators with 
arbitrarily many twist field insertions can be viewed as building blocks 
of the perturbation series of a relevant perturbation by the twist 
field. This ``tachyon condensation'' is responsible, e.g., for the 
formation of D0-D2 bound states, as discussed in \q{\GNS}. 
Upon integrating over twist field insertion points in the one-point 
functions $Z_{\Delta}(\vec  x)\,\exp\{ g^2 S_0(z)\}$, one would arrive 
at one-point functions which characterize the boundary theory after 
tachyon condensation. Sen's approach \q{\Sen} and the results of \q{\RStwo} 
allow one to circumvent the relevant boundary flow and to replace it 
by a combination of marginal bulk and boundary deformations. However, 
some questions as to the equivalence of both procedures remain open, 
and it might be useful to have an independent check of these methods. 
The correlation functions constructed here provide a starting point. 
\sn
For applications to superstring theory, it is mandatory to extend our 
analysis to free fermions. This does not pose serious problems, since 
systems of an even number of fermions can be bosonized. 
\sn
Problems of the type of our free boson problem are encountered 
in general boundary CFT as soon as the ``parent'' CFT on the plane 
admits different boundary conditions. For some general results on the 
rational case, see \q{\FSorbi,\FSgt,\FSms}. The spectrum of boundary 
condition changing operators can be derived as in Sect.\ 2, once 
boundary states for the ``constant'' boundary conditions are known. 
Again, the computation of correlators becomes non-trivial if boundary 
conditions with different gluing automorphisms are combined. 
In non-abelian WZW models, which constitute and important generalization 
of the free boson case, the Sugawara construction can be exploited in a 
similar fashion as for the free boson and leads to twisted, non-abelian 
Knizhnik-Zamolodchikov equations. The partition functions counting BCCOs 
in non-abelian boundary WZW models are linear combinations of the twining 
characters investigated in \q{\FSS} (see also \q{\BirFS} and references 
therein). Apart from the models with affine Lie algebra symmetry, there is 
the rather large class of so-called ``quasi-rational CFTs'' \q{\Na} on the 
plane for which generalizations of Knizhnik-Zamolodchikov equations 
exist even without a Sugawara form for the energy-momentum tensor 
\q{\ARS}. It might be interesting to see how such structures extend 
to boundary CFT. 

\bn
\subsection{Acknowledgments} We are indebted to P.\ Etingof for very 
helpful discussions. J.F.\ and O.G.\ thank C.\ Schweigert for 
interesting discussions. The work of O.G.\ is supported in part by the 
Department of Energy under Grant DE-FG02-94ER-25228 and by the National 
Science Foundation under Grant DMS-94-24334. O.G.\ also acknowledges the 
Clay Mathematics Institute for support.
\hbn
O.G.,\ A.R.\  and V.S.\ are grateful to the Research Group $M \cup \Phi$ 
at ETH Z\"urich for the warm hospitality extended to them.

\bn\bn   
\section{Appendix}
\vskip-.3cm
\subsection{A: Proofs of Lemma 1 and Lemma 2}%
\mn
{\sc Proof of Lemma 1:} We start from the usual expression for $\cT$ 
in  terms of $\cJ$ and rewrite it until we can perform the limit 
$\o_1  \rightarrow \o_2$. 
\ba{ \lefteqn{\cJ(\o_1) \, \cJ(\o_2) - \frac{1}{(\o_1-\o_2)^2}}
\nn \cr 
& =   \cJ_<(\o_1) \cJ(\o_2) + \cJ(\o_2) \cJ_>(\o_1) 
       + [\, \cJ_>(\o_1)\, ,\, \cJ(\o_2)\, ] - 
      \frac{1}{(\o_1-\o_2)^2} \nn \cr  
& =   \cJ_<(\o_1) \cJ(\o_2) + \cJ(\o_2) \cJ_>(\o_1) 
      + \frac{\frac12 \left(\frac{\o_1}{\o_2}\right)^{1/2} 
        + \frac{1}{2} \left( \frac{\o_2}{\o_1}\right)^{1/2} 
        - 1}{(\o_2 -\o_1)^2} \nn .
\cr }\ea 
We can now perform the limit $\o_1 \rightarrow \o_2=:\o$ to recover 
the generating field $\cT(\o)$ from the last formula 
$$ 
\cT(\o) =    \frac12\bigl(\, \cJ_<(\o) \cJ(\o) + \cJ(\o) 
             \cJ_>(\o) \,\bigr) + \frac{1}{16}\frac{1}{\o^2} 
$$
where \ we \ used \quad\quad${\displaystyle  \lim_{u\rightarrow 1}
     \frac{u^{1/2} + u^{-1/2} - 2}{2 (1-u)^2} \ = \ \frac{1}{8}\ }$.
\hfill\qed
\bn 

\noindent
{\sc Proof of Lemma 2:} The derivation of the commutation relation 
with $\cJ_>(\o)$ is straightforward. So, let us turn directly to the 
calculation of the commutator with $\cT_>(\o)$. Recall that $h=g^2/2$. Then, 
\def\nox{\left(\frac{i}{2z}\right)^h} 
\ba{ 
\lefteqn{[\, \cT_>(\o)\, ,\, \Psi_\rg (z)\,]}\nn\cr
& =   \nox \left(
\, [\cT_>(\o),e^{i\rg X_<(z)}]\, e^{i\rg X_>(z)} + 
 e^{i\rg X_<(z)}\, [\cT_>(\o),e^{i\rg X_>(z)}] \right) \nn \cr
& =   \nox e^{i\rg X_<(z)} \left(\, i \rg \, [\cT_>(\o),X_<(z)] - 
   h\, [X_<(x),[X_<(x),\cT_>(\o)]\,] \phantom{\frac12}\right. \nn \cr
 &   \hspace{1.2cm} +\left. \phantom{\frac12} i \rg \, [\cT_>(\o), X_>(z)] 
   + h\, [X_>(z),[X_>(z),\cT_>(\o)]\,]\right) e^{i\rg X_>(z)} \nn \cr
& =   \nox e^{i\rg X_<(z)} \left( \,i \rg \, [\cT_>(\o),X(z)] 
 + h\, (\sum_{r,s<0,n\geq-1} \o^{-n-2} z^n\, \d_{r+s,-n} \right. 
  \nn \cr 
 &   \hspace{4.3cm} \left. 
  - \sum_{r,s>0,n\geq -1} \o^{-n-2} z^n\, \d_{r+s,-n} )\right)
    e^{i\rg X_>(z)} \nn \cr
& =   \nox e^{i\rg X_<(z)} \left( 
  \, \frac{i\rg }{\o-z} \ \p_z X(z) + h\, \bigl(\sum_{n\geq 1} \ 
   \o^{-n-2} z^n \, n - \frac{1}{\o z}\bigr)\right) e^{i\rg X_>(z)}\nn 
  \cr
& =   \frac{1}{\o-z} \ \nox \p_z\left(\, e^{i\rg X_<(z)} 
 e^{i\rg X_>(z)}\right) + h\,\left( \frac{z}{\o}\frac{1}{(\o-z)^2}
 - \frac{1}{\o z} \right) \Psi_\rg (z) \nn \cr
& =   \frac{1}{\o-z}\ \p_z \Psi_\rg (z) + 
  \frac{1}{\o-z} \frac{h}{z} \Psi_\rg (z) + h\, \left(
  \frac{1}{(\o-z)^2} - \frac{1}{z(\o -z)}\right) \Psi_\rg (z) \nn 
  \cr
& =   \frac{1}{\o-z}\ \p_z\Psi_\rg (z) + \frac{h}{(\o-z)^2} \ 
    \Psi_\rg (z) \ \ . \nn 
\cr }\ea
In the process of this computation we have inserted the 
commutation relation between $L_n,a_r$ and eq.\ \refe{\TX}. 
The rest involves only standard algebraic manipulations. \hfill\qed      
\def\reg{{\rm reg}}
\medskip
\subsection{B: The function 
\hbox{\fat S}${}_{\hbox{\smfat 0}}$(\hbox{\fat z})}%
\mn
In this appendix we want to explain a number of properties of 
the function $S_0(z)$ that is introduced in Section 4.1. To show
that the limit $\lim_{v \rightarrow x}$ exists, we insert the 
definition of $\lambda_0(\xi)$ into eq.\ \Sfun\ . After 
splitting off all non-singular terms in $\lambda_0$ we obtain: 
\ba{  
S_0(z) &\ = \   \lim_{v \rightarrow x}\left[^{\phantom{\frac12}} 
     2 \ \Re \int_v^z \ \lambda_0(\xi) \, d\xi - \log(v-\bv) - 
     \Re \log P(v) \ \right]\ \bcr
&\ = \ \lim_{v \rightarrow x}\left[^{\phantom{\frac12}} 
     - 2 \ \Re \int_v^z \ \frac{1}{\xi-\bar \xi} \, d\xi  - \log(v-\bv)  
     \  + \reg_{v \rightarrow x}\; \right]\ \bcr 
& \ = \ \lim_{v \rightarrow x} \left[\ \log (v-\bv) - \log(z-\bz) - 
    \log(v-\bv)  + {\reg'}_{v \rightarrow x} \ \right]\ \ . 
}\ea 
Since the singularity from the integral cancels against the 
term $\log(v - \bv)$, the limit can be taken. 
\mn 
Our second aim is to understand that $S_0(z) \equiv S^x_0(z)$ does not 
depend on the choice of $x$. Let us displace $x$ by some small amount 
$a \in \R$ such that $x + a$ is still in the Dirichlet 
interval $D_1$. Comparison of $S_0^x(z)$ and $S^{x+a}_0(z)$ gives
\ba{S_0^x(z) & - S^{x+a}_0(z) \bcr  & 
    \ = \ \lim_{v \rightarrow x}
        \left[^{\phantom{\frac12}} 
     2 \ \Re \int_v^{v+a} \ \lambda_0(\xi) \, d\xi - 
     \Re \log P(v) \ + \Re \log P(v+a)\ \right]\bcr 
& \ = \  \lim_{v \rightarrow x} \left[\, - 2 \Re  \int_v^{v+a}
    \sum_{i=1}^{2g-1} \frac{\frac12}{\xi-x_i} \, d\xi \ + \ 
  \Re \log \frac{P(v + a)}{P(v)}\; \right] \bcr 
& \ = \  \lim_{v \rightarrow x}\left[\, - \Re \sum_{i=1}^{2g+1} 
    \, \left( \log(v+a-x_i) - \log(v-x_i) \right) \ +\        
   \Re \log \frac{P(v + a)}{P(v)} \;\right] \ = \ 0 \ \ . \cr 
} \ea 
In passing to the second line we omitted all terms in the integrand
which vanish when $\xi$ comes close to the real axis. 
\mn
Finally, we investigate the behaviour of $S_0(z)$ at the boundary. 
Basically, one repeats the analysis we have sketched above in our 
discussion of $\lim_{v \rightarrow x}$. If the end-point $z$ of our 
integration approaches one of the Dirichlet intervals, this leads to 
the singularity $ \sim - \log|z-\bz|$. In the argument one needs that 
the quotient $\sqrt{P(z)/P(\bz)}$ in front of the singular term 
$1/(z-\bz)$ satisfies $\lim_{z \rightarrow x} \sqrt{P(z)/P(\bz)} = 1$ 
for $x \in D_k$. This is no longer true when $z$ is sent to the real 
axis in one of the Neumann intervals $N_k$. In fact, upon moving $x$ 
from a Neumann into a Dirichlet interval, the polynomial $P(x)$ changes 
sign, causing the quotient $P(z)/P(\bz)$ to surround the origin of the 
complex plane once. After taking the square root we conclude that 
$\lim_{z \rightarrow x} \sqrt{P(z)/P(\bz)} = -1 $ for $x \in N_k$ and 
hence $S_0(x) \sim \log|z - \bz|$ near the Neumann intervals.    
\bn\bn\bn\bn
\def\q#1{\cr$\lb{\rm #1}\rb$}
\vfill\eject
\parindent=35pt \vsize=23.3truecm
\noindent
\halign{#\phantom{t}\hfil&\vtop{\parindent=0pt \hsize=37.3em #\strut}\cr
\noalign{\leftline{{\large References }}} \noalign{\vskip.4cm}
$\lb{\AfL}\rb$ &I.\  Affleck, A.W.W.\  Ludwig, {\sl  
   Critical theory of overscreened Kondo fixed points},  
   Nucl.\ Phys.\ B{\bf360} (1991) 641
\q{\ARS} &A.Yu.\ Alekseev, A.\ Recknagel, V.\ Schomerus, 
     {\sl Generalization of the Knizhnik-Zamo\-lodchikov equations}, 
     Lett.\ Math.\ Phys.\ {\bf41} (1997) 169, hep-th/9610066
\q{\ARStwo} &A.Yu.\ Alekseev, A.\ Recknagel, V.\ Schomerus, 
     {\sl Non-commutative world-volume geometries: branes on SU(2) and 
    fuzzy spheres}, J.\ High Energy Phys.\ 9909 (1999) 023, hep-th/9908040
\q{\ABMNV} &L.\ Alvarez-Gaume, J.B.\ Bost, G.\ Moore, P.\ Nelson, 
   C.\ Vafa, {\sl Bosonization on higher genus Riemann surfaces},
  Commun.\ Math.\ Phys.\ {\bf 112} (1987) 503
\q{\BirFS} &L.\ Birke, J.\ Fuchs, C.\ Schweigert, {\it Symmetry 
     breaking boundary conditions and WZW orbifolds}, hep-th/9905038 
\q{\BDLR} & I.\ Brunner, M.R.\ Douglas, A.\ Lawrence, C.\ R\"omelsberger, 
  {\sl D-branes on the quintic}, hep-th/9906200
\q{\CLNY} &C.G.\ Callan, C.\ Lovelace, C.R.\ Nappi, S.A.\ Yost, 
 {\sl Adding holes and crosscaps to the superstring}, Nucl.\ Phys.\ 
 B{\bf293} (1987) 83; \quad {\sl Loop corrections to superstring equations 
 of motion},  Nucl.\ Phys.\ B{\bf308} (1988) 221
\q{\Carold}&J.L.\ Cardy, {\sl Conformal invariance and surface
 critical behavior}, Nucl.\ Phys.\ B{\bf240} (1984) 514; 
  \quad {\sl Effect of boundary conditions on the
 operator content of two-dimensional conformally invariant theories},
 Nucl.\ Phys.\ B{\bf275} (1986) 200
\q{\Carfus}&J.L.\ Cardy, {\sl Boundary conditions, fusion rules
 and the Verlinde formula}, Nucl.\ Phys.\ B{\bf324} (1989) 581
\q{\CarLew} &J.L.\ Cardy, D.C.\ Lewellen, {\sl Bulk and boundary operators
 in conformal field theory}, Phys.\ Lett.\ B{\bf259} (1991) 274
\q{\CIMM}&B.\ Chen, H.\ Itoyama, T.\ Matsuo, K.\ Murakami, {\sl p-p' 
 system with B-field, branes at angles and noncommutative geometry},
 hep-th/9910263
\q{\CorFai} &E.\ Corrigan, D.B.\ Fairlie, {\sl Off-shell states in dual 
   resonance theory}, Nucl.\ Phys.\ B{\bf91} (1975) 527
\q{\DiaRoem} &D.-E.\ Diaconescu, C.\ R\"omelsberger, {\sl D-branes and 
   bundles on elliptic fibrations}, hep-th/9910172
\q{\Dou}&M.R.\ Douglas, {\sl Gauge Fields and D-branes}, J.\ Geom.\ 
 Phys.\ {\bf28} (1998) 255, hep-th/9604198
\q{\DH} &M.R.\ Douglas, C.\ Hull, {\sl D-branes and the noncommutative 
  torus}, J.\ High Energy Phys.\ 9802 (1998) 008, hep-th/9711165
\q{\FFFSone} & G.\ Felder, J.\ Fr\"ohlich, J.\ Fuchs, C.\ Schweigert, 
   {\sl The geometry of WZW branes}, hep-th/9909030
\q{\FM} &J.\ Fr\"ohlich, P.A.\ Marchetti, {\sl Bosonization, topological
solitons and fractional charges in two dimensional quantum field theory}, 
Commun.\ Math.\ Phys.\ {\bf116} (1988) 127
\q{\FSS} &J.~Fuchs, B.~Schellekens, C.~Schweigert,
  {\sl From Dynkin diagram symmetries to fixed point structures}, 
  Commun.\ Math.\ Phys.\ {\bf180} (1996) 39, 
 hep-th/9506135; \quad  
 {\sl Twining characters, orbit Lie algebras, and fixed point resolution},
  q-alg/9511026
\q{\FS} &J.\ Fuchs, C.\ Schweigert, {\sl Branes:\ from free fields to general 
  conformal field theories}, Nucl.\ Phys.\ B{\bf530} (1998) 99, 
  hep-th/9712257
\q{\FSorbi} &J.\ Fuchs, C.\ Schweigert, {\sl Orbifold analysis of broken bulk symmetries}, hep-th/9811211
\q{\FSgt} &J.\ Fuchs, C.\ Schweigert, {\sl Symmetry breaking boundaries I. General theory}, hep-th/ 9902132
\q{\FSms} &J.\ Fuchs, C.\ Schweigert, {\sl Symmetry breaking boundaries II. More structures; examples}, hep-th/9908025
\q{\GNS} &E.\ Gava, K.S.\ Narain, M.H.\ Sarmadi, {\sl On the bound states 
  of p- and (p+2)-branes}, Nucl.\ Phys.\ B{\bf504} (1997) 214, hep-th/9704006
\q{\GuS} &M.\ Gutperle, Y.\ Satoh, {\sl D-branes in Gepner models 
  and supersymmetry}, Nucl.\ Phys.\  B{\bf543} (1999) 73, 
  hep-th/9808080; \quad  {\sl D0-branes in Gepner models and 
  \hbox{$N=2$} black holes}, hep-th/9902120
\q{\Hash} &A.\ Hashimoto, {\sl Dynamics of Dirichlet-Neumann open strings 
   on D-branes}, Nucl.\ Phys.\  B{\bf496} (1997) 243, hep-th/9608127
\q{\Ish} &N.\ Ishibashi, {\sl The boundary and crosscap states
 in conformal field theories}, Mod.\ Phys.\ Lett.\ A{\bf4} (1989) 251
\q{\Mal} &J.M.\ Maldacena, {\sl Black holes in string theory}, hep-th/9607235
\q{\Na} &W.\ Nahm, {\sl Quasi-rational fusion products}, 
  Int.\ J.\ Mod.\ Phys.\ B{\bf8} (1994) 3693, hep-th/9402039 
\q{\OOY} &H.\ Ooguri, Y.\ Oz, Z.\ Yin, {\sl D-branes on
 Calabi-Yau spaces and their mirrors}, Nucl.\ Phys.\ B{\bf477}
 (1996) 407, hep-th/9606112
\q{\Pol} &J.\ Polchinski, {\sl Dirichlet branes and Ramond-Ramond 
 charges}, Phys.\ Rev.\ Lett.\ {\bf 75} (1995) 4724, hep-th/9510017; \quad 
   {\sl TASI lectures on D-branes}, hep-th/9611050
\q{\Rai} &A.K.\ Raina, {\sl An algebraic geometry study of the b-c 
   system with arbitrary twist fields and arbitrary statistics}, 
   Commun.\ Math.\ Phys.\ {\bf140} (1991) 373
\q{\RSone} &A.\ Recknagel, V.\ Schomerus, {\sl D-branes in Gepner models}, 
 Nucl.\ Phys.\ B{\bf531} (1998) 185, hep-th/9712186
\q{\RStwo} &A.\ Recknagel, V.\ Schomerus, 
     {\sl Boundary deformation theory and moduli spaces of D-branes}, 
     Nucl.\ Phys.\ B{\bf545} (1999) 233, hep-th/9811237
\q{\Sag} &A.\ Sagnotti, {\sl Open strings and their symmetry groups}, in:\   
  Non-Perturbative Methods in Field Theory, eds.\ G.\ Mack et al., 
  Lecture Notes Carg\`ese 1987;\ \   
   {\sl Surprises in open-string perturbation theory}, 
   Nucl.\ Phys.\ Proc.\ Suppl.\ {\bf56B} (1997) 332, hep-th/9702093%
\q{\Vol} &V.\ Schomerus,  {\sl D-branes and deformation quantization}, 
    J.\ High Energy Phys.\ 9906 (1999) 030, hep-th/9903205
\q{\SW} &N.\ Seiberg, E.\ Witten, {\sl String theory and noncommutative 
  geometry}, J.\ High Energy Phys.\ 9909 (1999) 032, hep-th/9908142
\q{\Sen}&A.\ Sen, {\sl SO(32) spinors of type I and other solitons on 
   brane-antibrane pair}, J.\ High Energy Phys.\ 9809 (1998) 023, 
  hep-th/9808141; \ \ {\sl Descent relations among bosonic D-branes}, 
   hep-th/9902105; \ \ {\sl Non-BPS states and branes in string theory}, 
   hep-th/9904207%
\q{\Sta}&S.\ Stanciu, {\sl D-branes in Kazama-Suzuki models}, Nucl.\ Phys.\  
   B{\bf 526} (1998) 295, hep-th/9708166
\q{\SV}&A.\ Strominger, C.\ Vafa, {\sl Microscopic origin of the 
  Bekenstein-Hawking entropy}, Phys.\ Lett.\ B{\bf379} (1996) 99, 
  hep-th/9601029
\q{\Wibou}&E.\ Witten, {\sl Bound states of strings and D-branes}, 
 Nucl.\ Phys.\ B{\bf460} (1996) 335, hep-th/9510135
\q{\WiK}&E.\ Witten, {\sl D-branes and K-theory}, J.\ High Energy Phys.\ 
  9812 (1998) 019, hep-th/9810188
\q{\Zamo} &Al.B.\ Zamolodchikov, {\sl Conformal scalar field on the 
  hyperelliptic curve and critical Ashkin-Teller multipoint correlation 
  functions}, Nucl.\ Phys.\ B{\bf285} (1987) 481
\cr}

\bye